\documentclass[aps,pre,noeprint,twocolumn,amssymb,superscriptaddress,longbibliography]{revtex4-1}
\usepackage{amsmath}
\usepackage{amsfonts}
\usepackage{amssymb}
\usepackage{epsfig}
\usepackage{graphicx}
\usepackage[normalem]{ulem}

\renewcommand{\phi}{\varphi}

\newcommand{\be}{\begin{equation}}
\newcommand{\ee}{\end{equation}}
\newcommand{\bea}{\begin{equnaray}}
\newcommand{\eea}{\end{equnaray}}
\newcommand{\ba}{\begin{align}}
\newcommand{\ea}{\end{align}}

\usepackage{color}

\definecolor{green}{rgb}{0.0, 0.44, 0.0}
\definecolor{red}{rgb}{1.0, 0.13, 0.32}
\definecolor{blue}{rgb}{0.06, 0.2, 0.65}
\definecolor{magenta}{rgb}{1.0, 0.0, 1.00}
\definecolor{purple}{rgb}{0.7, 0.0, 0.7}
\definecolor{cyan}{rgb}{0.0, 1.0, 1.0}

\begin{document}

\title{Rare events and disorder control the brittle yielding of well-annealed amorphous solids}

\author{Misaki Ozawa}

\affiliation{Laboratoire de Physique de l'Ecole normale sup\'erieure, ENS, Universit\'e PSL, CNRS, Sorbonne Universit\'e, Universit\'e Paris-Diderot, Sorbonne Paris Cit\'e, 75005 Paris, France}

\author{Ludovic Berthier}
\affiliation{Laboratoire Charles Coulomb (L2C), Universit\'e de Montpellier, CNRS, 34095 Montpellier, France}
\affiliation{Department of Chemistry, University of Cambridge, Lensfield Road, Cambridge CB2 1EW, United Kingdom}

\author{Giulio Biroli}
\affiliation{Laboratoire de Physique de l'Ecole normale sup\'erieure, ENS, Universit\'e PSL, CNRS, Sorbonne Universit\'e, Universit\'e Paris-Diderot, Sorbonne Paris Cit\'e, Paris, France}

\author{Gilles Tarjus}
\affiliation{LPTMC, CNRS-UMR 7600, Sorbonne Universit\'e, 4 Place Jussieu, 75252 Paris cedex 05, France}

\begin{abstract}
We use atomistic computer simulations to provide a microscopic description of the brittle failure of amorphous materials,  and we assess the role of rare events and quenched disorder. We argue that brittle yielding originates at rare soft regions, similarly to Griffiths effects in disordered systems. We numerically demonstrate how localized plastic events in such soft regions trigger macroscopic failure via the propagation of a shear band. This physical picture, which no longer holds in poorly annealed ductile materials, allows us to discuss the role of finite-size effects in brittle yielding and reinforces the similarities between yielding and other disorder-controlled nonequilibrium phase transitions. 
\end{abstract}

\maketitle

\section{Introduction}

Yielding of slowly deformed amorphous solids can proceed in two qualitatively different ways. Whereas ductile materials reach a stationary plastic flow through a continuous evolution under applied deformation, brittle ones undergo a macroscopic failure at which the stress discontinuously drops via the formation of a system-spanning 
shear band~\cite{nicolas2018deformation,barrat2011heterogeneities,rodney2011modeling,bonn2017yield,falk2011deformation}. We recently argued that the two regimes can be observed in the same material if prepared over a wide enough range of annealing conditions, and they are representative of distinct phases of yielding, separated by a critical point reminiscent of that found in an  out-of-equilibrium random-field Ising model (RFIM)~\cite{ozawa2018random,singh2020brittle,ozawa2020role}.

Here, we provide a microscopic perspective on the brittle yielding of amorphous media, shedding light on two issues that are central to understand material failure. First, we focus on the microscopic origin of shear banding. This is a widely studied question in the context of metallic glasses where it is of great practical interest for improving their ductility~\cite{greer2013shear}. Many studies on metallic glasses have considered in particular the differing  characteristics of homogeneous and heterogeneous shear-band nucleation~\cite{cheng2011intrinsic,tian2012approaching}, the former being a property of an ideal bulk material whereas the latter is mainly due to extrinsic flaws such as impurities or surface imperfections. Second, we take a new angle to investigate how finite-size effects modify brittle yielding. This is again relevant for experiments on metallic glasses in relation with a possible evolution towards brittleness~\cite{yang2012size,wang2016sample} and is also central for the theory of yielding. In fact, the existence of a critical point as a function of glass stability has recently been challenged~\cite{barlowductile}, and it was attributed to putative finite-size effects beyond those already analysed in Refs.~ \cite{ozawa2018random,ozawa2020role}. 
The sudden macroscopic failure via shear-band formation studied here differs from the shear banding discussed in the steady-state deformation of ductile materials~\cite{amitrano2006brittle,vandembroucq2011mechanical,nicolas2014universal} but may be relevant for oscillatory deformation in brittle materials~\cite{yeh2020glass}.

The crux of the present study stems from analogous work performed in the context of nonequilibrium transitions in disordered systems such as the driven RFIM at zero temperature~\cite{sethna2006random}. We demonstrate that similarly to the zero-temperature spinodal of the RFIM~\cite{nandi2016spinodals}, brittle yielding in well-annealed amorphous solids is controlled by rare events and quenched disorder~\cite{nandi2016spinodals,popovic2018elastoplastic,rossi2022emergence}. Disorder obviously refers to the amorphous structure of the material~\cite{rossi2022emergence} but the nature of "rare events" is more subtle and is worth discussing first. It is well established~\cite{argon1979plastic,falk1998dynamics,maloney2006amorphous,nicolas2018deformation,dasgupta2013yield} that slow deformation of amorphous solids involves localized plastic events that are characterized by local-stress and nonaffine-displacement fields having a quadrupolar symmetry similar to that of Eshelby inclusions in continuum elasticity theory~\cite{eshelby1957determination}. Furthermore, shear-band formation seems to be associated with a mechanical instability taking the form of a line  (in $2 d$) or a plane (in $3 d$) of Eshelby quadrupoles~\cite{dasgupta2013yield,dasgupta2013shear,csopu2017atomic,hieronymus2017shear,hassani2019probing}. The structural origin of soft regions
prone to plastic rearrangements has been studied intensively (see Ref.~\cite{richard2020predicting} for a critical review). As confirmed by numerical simulations, plastic events generically occur at ``weak spots''. Crucially, {\it large} weak spots become very rare in stable glassy materials. To understand this, recall that correlation lengths are rather modest 
in supercooled liquids near the glass transition temperature~\cite{berthier2011theoretical}. Let $\xi$ be the typical glassy length and $p$ the probability to find a soft poorly-annealed region of volume $\xi^d$, where $d$ is the spacial dimension, around a given point in a stable glass. The probability to find a much larger soft region of volume $v$ is proportional to $p^{v/\xi^d}\sim \exp(-c v/\xi^d)$ with $c$ is a constant of order unity. This exponential suppression, characteristic of Griffiths phases 
in disordered media~\cite{griffiths1969nonanalytic}, statistically makes large soft regions a rare occurrence in well-annealed glasses. 

Computer simulations are therefore unable to directly probe such a phenomenon, as only small soft regions are found even in the largest systems that can be studied numerically. 
As a result, the central role of rare events is completely missed by the simulations. To get around this major difficulty, we insert a soft region in an otherwise stable glass by fiat~\cite{ozawa2018random,popovic2018elastoplastic}. Knowing that shear bands extend preferentially along the direction of shear, we choose an elongated shape oriented along the expected direction of shear banding. (We discuss the role of the seed anisotropy in Appendix \ref{app:location}). The key physical point is that in a well-annealed amorphous material such a rare soft region is more prone to rearrange under shear deformation and provides a ``seed'' for shear bands. Under applied deformation this region yields early, relaxing stress before the bulk of the material. By doing so, it destabilizes surrounding particles, which as a consequence also yield before the bulk. This leads to an extension of the soft region along the principal 
direction of the seed and, for a certain value of the imposed shear strain, to a self-sustained process leading to macroscopic shear banding. 

Our numerical procedure mimics the spontaneous nucleation of a single shear-band embryo~\cite{wang2016sample} at a single intrinsic soft defect in a pristine {\it macroscopic} sample that we cannot simulate directly. The strong finite-size effect due to the difficulty in finding exponentially rare defects is thus fully circumvented. The aim of this work is to unveil and study the microscopic origin of failure for brittle yielding. Given that defects are very rare, their interaction is not expected to play any role in triggering the instability,  in contrast to other materials, where local damages percolate~\cite{weiss2014finite,shekhawat2013damage}. A complete study of the entire process of brittle yielding also needs to take into account how the shear bands triggered by very distant defects interact. This is beyond the scope of this work and is left for future studies.  

The defects considered in this work are not extrinsic defects of the type that has been implemented in coarse-grained models, such as surface imperfections, notches~\cite{rosti2001pinning,vasoya2016notch}, or hard inclusions~\cite{tyukodi2016finite}. Despite the fact that we add the seeds by hand, seeds represent rare intrinsic defects associated with weaker particle arrangements within the bulk material: they spontaneously exist in macroscopic samples but are rare due to their low probability, contrary to extrinsic ones which would not exist in pristine samples. As a consequence, the phenomenon that we study differs from heterogeneous nucleation. It also differs from the body of work that considers the effect of a preexisting crack on the fracture of a material. The defects here are not voids but soft regions that can undergo multiple plastic rearrangements and are necessarily present as spontaneous structural fluctuations in a well-annealed but very large amorphous solid.

\begin{figure}
\includegraphics[width=0.95\columnwidth]{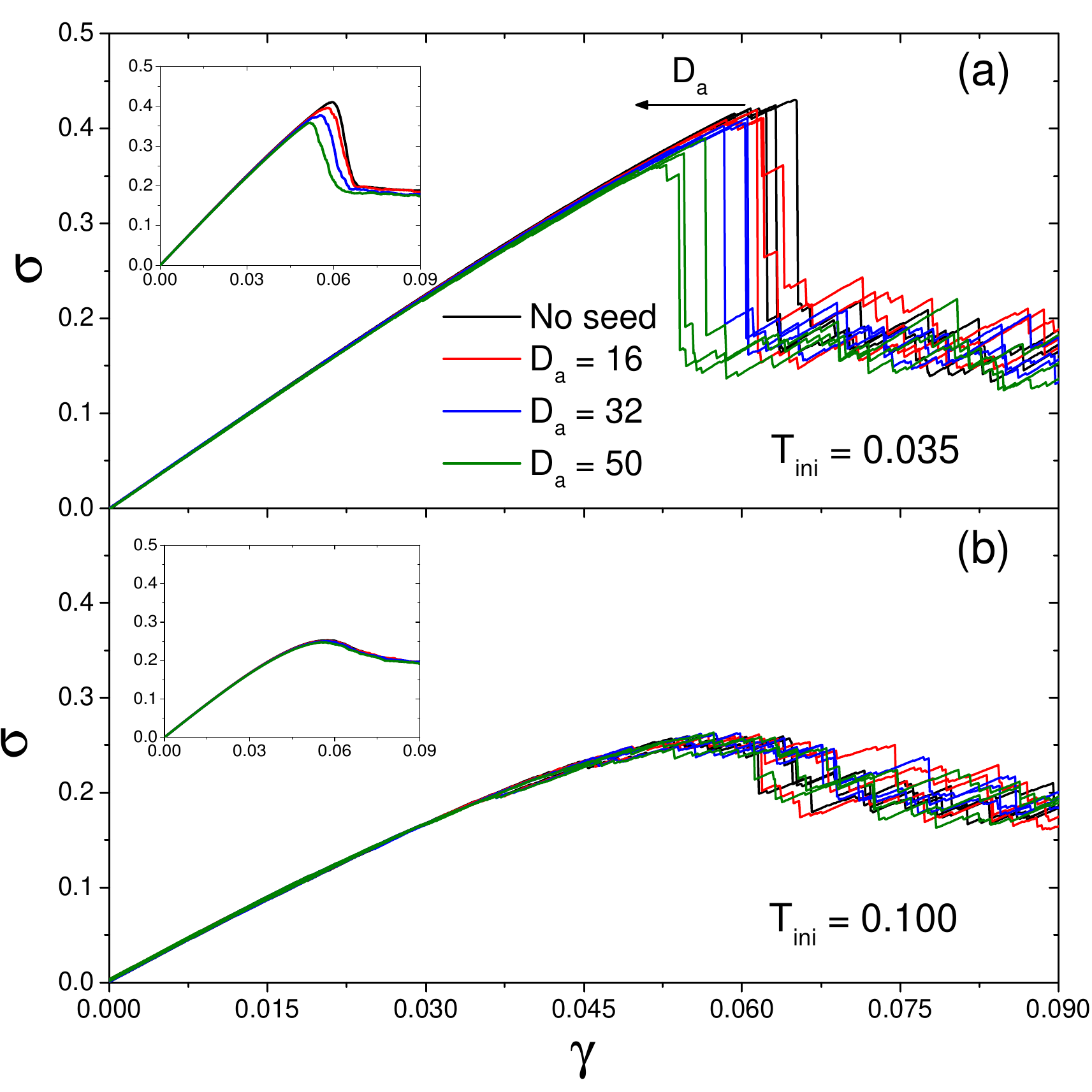}
\includegraphics[width=0.95\columnwidth]{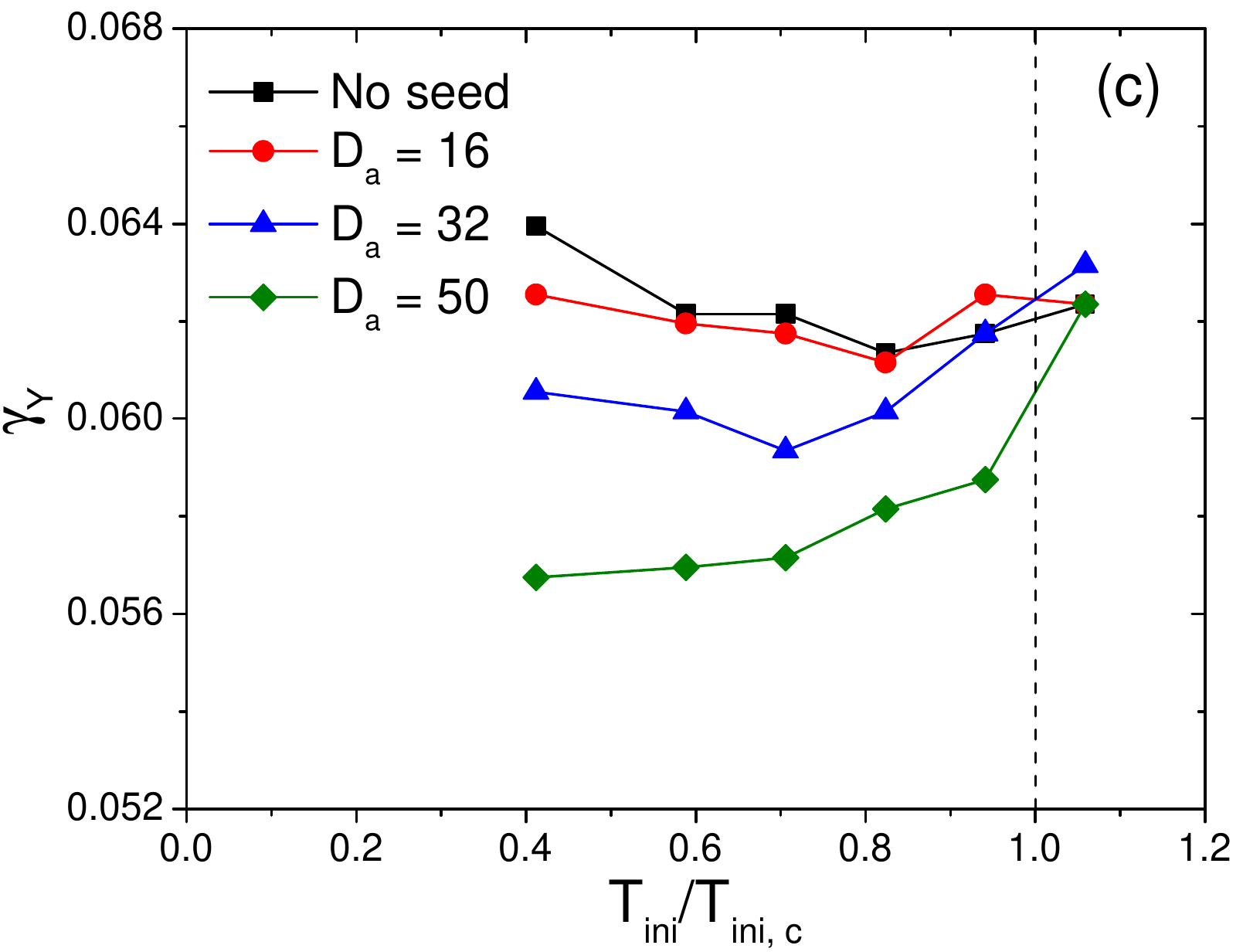}
\caption{Effect of an elongated {soft} seed of size $D_a\times D_b$ on the stress versus strain curves of $2d$ glass samples with $N=64000$ atoms and fixed 
$D_b=8$.
(a,b) Three independent realisations for each $D_a$ are shown for a stable glass with $T_{\rm ini}=0.035$ (a) and a poorly annealed glass with $T_{\rm ini}=0.100$ (b). 
Insets show the average over $100-400$ samples. (c) Yield strain $\gamma_{\rm Y}$ versus glass fictive temperature $T_{\rm ini}$ (normalized by the 
{apparent} critical value $T_{\rm c}=0.085$ for the $N=64000$ system) for several seed lengths.}
\label{fig:stress}
\end{figure} 

\begin{figure*}[t!]
\includegraphics[width=0.66\columnwidth]{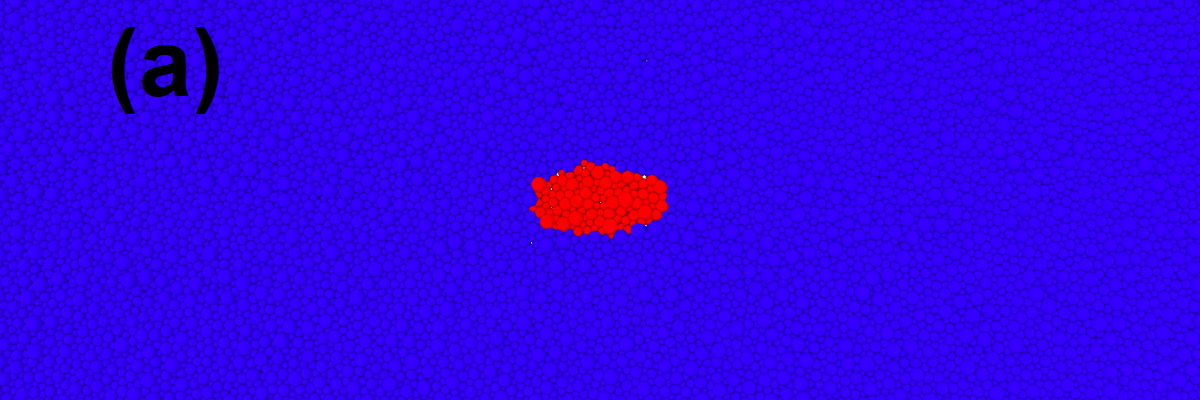}
\includegraphics[width=0.66\columnwidth]{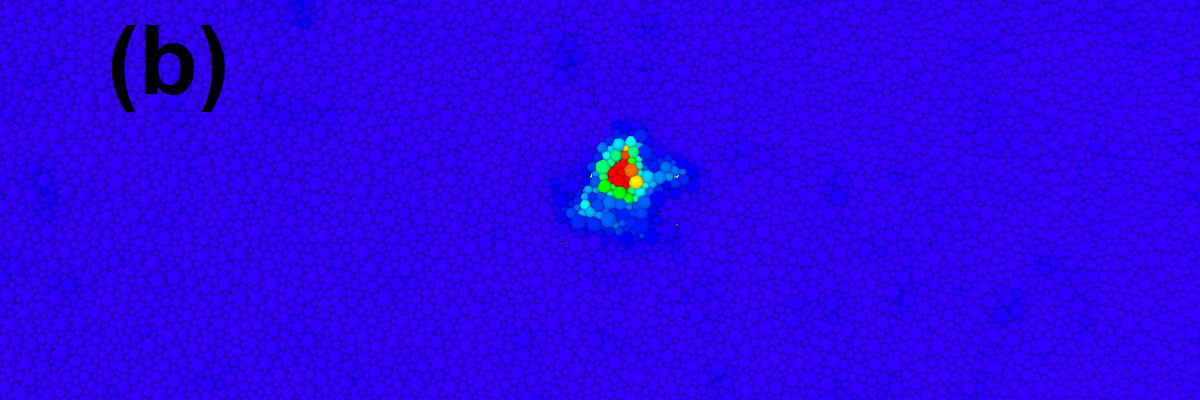}
\includegraphics[width=0.66\columnwidth]{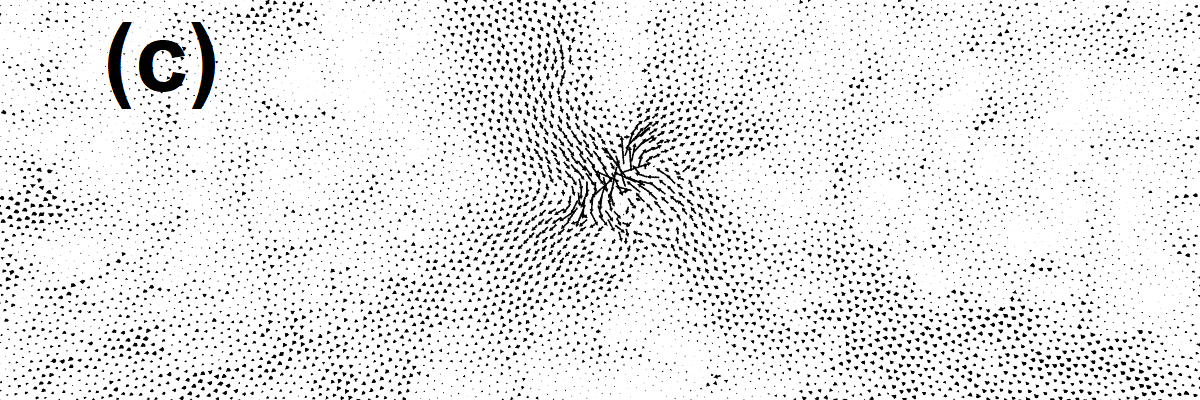}
\includegraphics[width=0.66\columnwidth]{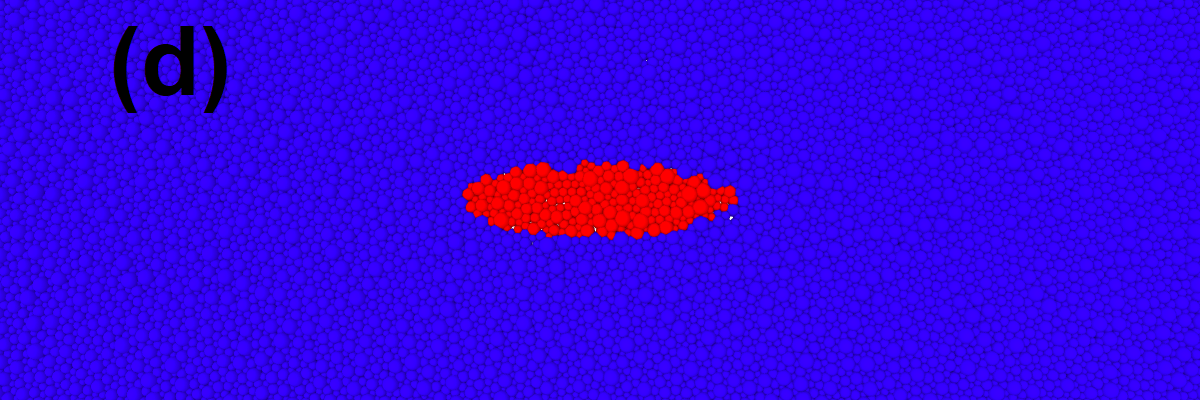}
\includegraphics[width=0.66\columnwidth]{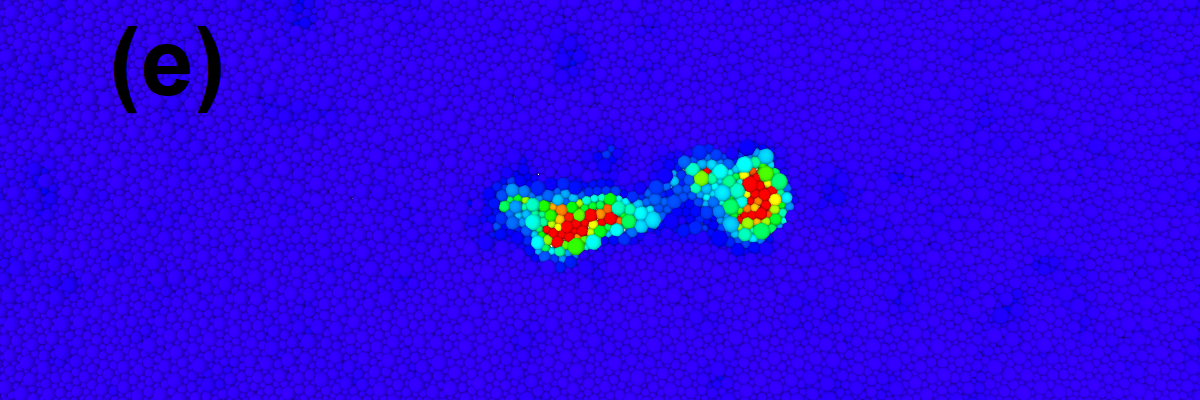}
\includegraphics[width=0.66\columnwidth]{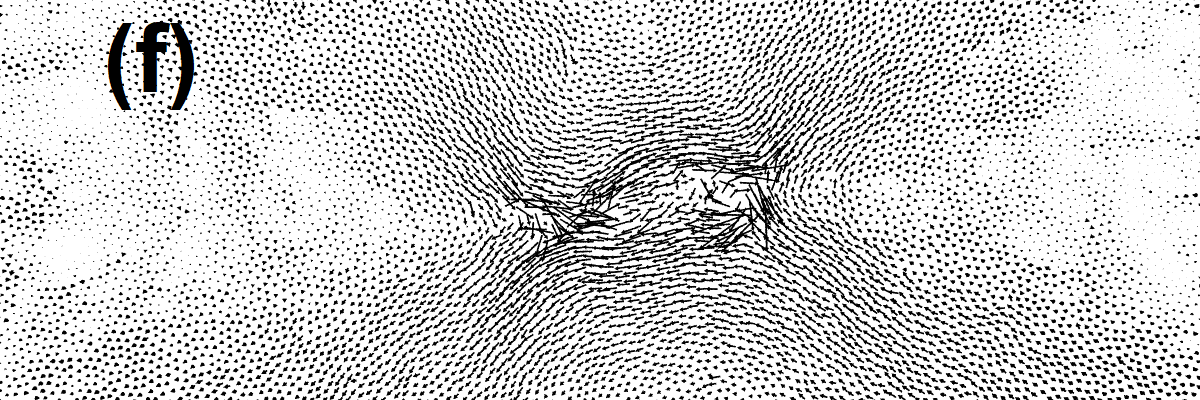}
\includegraphics[width=0.66\columnwidth]{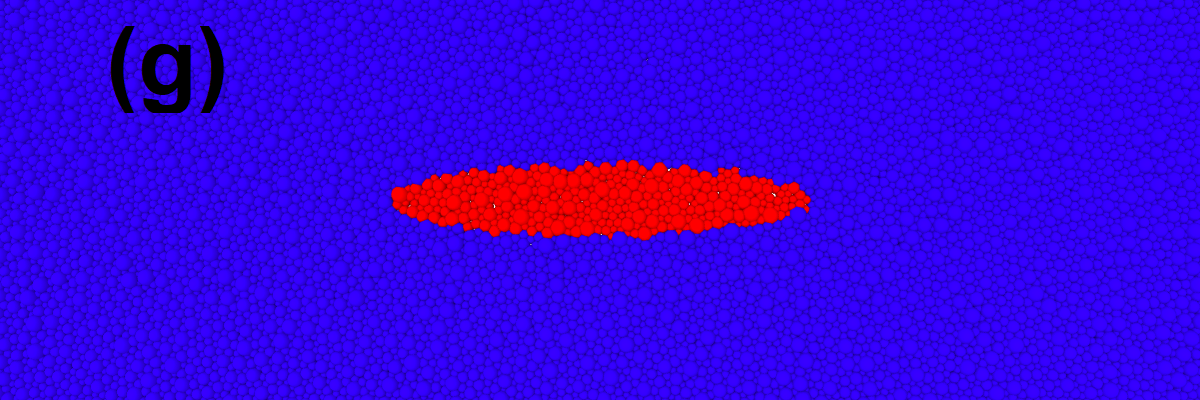}
\includegraphics[width=0.66\columnwidth]{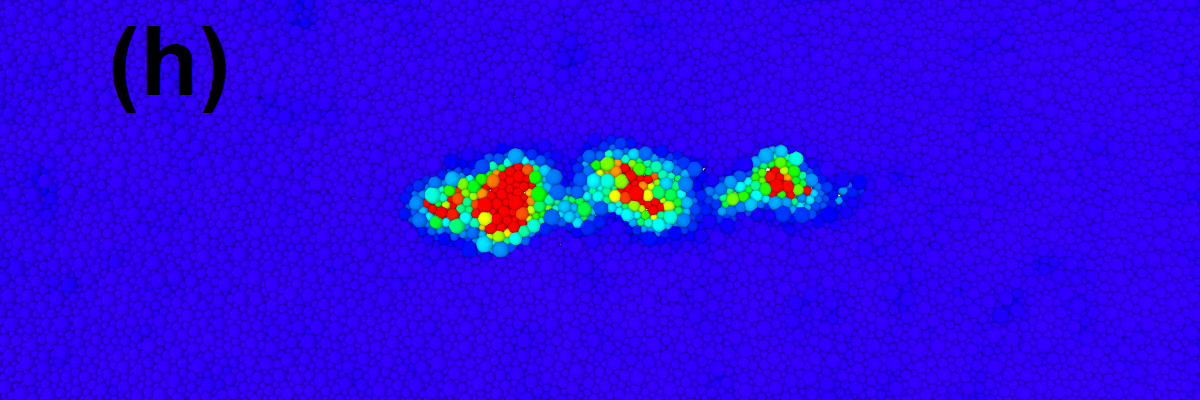}
\includegraphics[width=0.66\columnwidth]{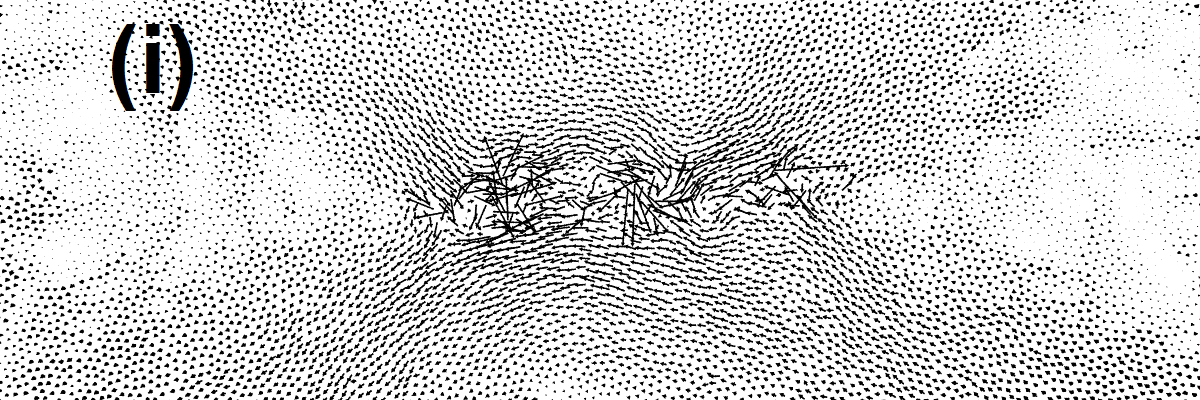}
\caption{Visualization of the seed region for a stable glass with $T_{\rm ini}=0.035$. Left panels (a,d,g): Particles inside the seed and in the rest of the sample are 
colored in red and blue, respectively.
Middle panels (b,e,h): Nonaffine square displacement $D_{\rm min}^2$ between the initial configuration and the configuration in the elastic-like regime at $\gamma=0.03$. 
Right panels (c,f,i): Nonaffine displacement vectors (magnified by a factor 10) corresponding to the middle panels. Top (a,b,c): $D_a=16$. Middle (d,e,f): $D_a=32$. 
Bottom(g,h,i): $D_a=50$.}
\label{fig:before_yielding}
\end{figure*}

\section{Methods}

For simplicity, we focus on the simple shear of a $2d$ glass since the brittle-to-ductile transition and the analogy with the RFIM hold in $2 d$~\cite{ozawa2020role}. In addition, to be able to make crisp statements about nonequilibrium phase transitions, instabilities and other singularities, we consider the limit of zero temperature and of a quasi-static applied strain. Our study is based on the simulation of a $2d$ glass composed of polydisperse soft disks~\cite{berthier2019zero}. We prepare equilibrium 
supercooled liquid configurations over a very wide range of temperatures $T_{\rm ini}$ by means of optimized swap Monte-Carlo simulations~\cite{ninarello2017models}. A ``seed'' is then inserted as follows. We define an ellipsoidal region characterized by $D_a$, the length of the major axis chosen along the direction of shear, and by 
$D_b$, the length of the minor axis. In a $2d$ system of $N$ atoms with $N$ from $8000$ to $64000$, which corresponds to a linear size $L$ from $89.4$ to $253.0$ (in units of the average atomic diameter), we fix $D_b=8$, which is of the order of the typical scale of an elementary rearranging region~\cite{barbot2018local}, and we vary $D_a$ up to $D_a=50$ which is of the order of half the smallest system size (see below for a discussion). We then perform additional Monte-Carlo simulations inside the seed region at a very high temperature $T_{\rm h}=10$ (about $100$ times larger than the mode-coupling crossover~\cite{berthier2019zero}). We quench the obtained atomic configuration to zero temperature by using the conjugate gradient method. As a result, the glass sample contains a poorly-annealed elliptic seed of length  $D_a$ inside an otherwise very stable material of linear size $L$. Our goal is to vary $D_a$ and $L$ systematically to infer the behavior of the system in the limit $L \gg D_a \gg 1$ in order to understand how a single shear band is initiated by a single rare region in a macroscopic sample.

Finally, we deform the samples by athermal quasi-static shear simulations at zero temperature with Lees-Edwards periodic boundary conditions, which correspond to change the shear strain in a controlled way throughout the sample~\cite{maloney2006amorphous}. More details on the model and methods are given in Appendix~\ref{sec:method_SI}.

\section{Results}

We start by establishing that rare seeds are essential in well-annealed glasses displaying brittle yielding. We first show stress $\sigma$ versus strain $\gamma$ curves 
for stable glasses in Fig.~\ref{fig:stress}(a). The curves with no seed show a large stress overshoot and a large  abrupt stress drop, as found previously~\cite{ozawa2018random,ozawa2020role}. When inserting a seed in the same initial configurations, the elastic-like regime of the $\sigma$ vs $\gamma$ curves 
seems barely modified. When the strain $\gamma$ is further increased, the samples with a seed continue to yield abruptly, but they do so systematically earlier for longer 
seeds. This directly establishes that a single soft region, which corresponds to a rare event in a pristine macroscopic glass, has a dramatic impact on bulk yielding.
We confirm this central fact also for seeds with different values of $D_b$ in Appendix~\ref{app:db}.

By contrast, the  stress versus strain curves for a poorly-annealed glass with $T_{\rm ini}=0.100$ in Fig.~\ref{fig:stress}(b) hardly show any change with the presence or the length of the seed, suggesting that yielding is then not affected by rare events. We quantitatively support this conclusion by locating the average yield strain $\gamma_{\rm Y}$ for each degree of annealing (measured by $T_{\rm ini}$) and each $D_a$. We do so by using the maximum of the 
disconnected susceptibility introduced in Ref.~\cite{ozawa2018random} (see Appendix~\ref{app:gammay}). The systematic evolution of $\gamma_{\rm Y}$ is summarized in Fig.~\ref{fig:stress}(c) for $N=64000$. 
This plot is reminiscent of the RFIM study presented in Fig. 1 of Ref.~\cite{nandi2016spinodals}, which shows a phase diagram in the plane defined by the disorder strength $R$ and the applied coercive field $H_{\rm c}$ plane. $R$ and $H_{\rm c}$ in the RFIM play the same role as $T_{\rm ini}/T_{\rm c}$ and $\gamma_{\rm Y}$ in Fig.~\ref{fig:stress}(c), respectively.
It shows that the effect of seed insertion on the yielding transition fades away in the vicinity of the apparent critical point that we have previously located for $N=64000$~\cite{ozawa2020role}, akin to Fig. 1 of Ref.~\cite{nandi2016spinodals}.

Having established in which regime rare events dominate yielding, we now focus on the microscopic behavior giving rise to the observed seed influence. The snapshots in Fig.~\ref{fig:before_yielding} confirm that the tiny decrease of $\sigma$ before yielding ($\gamma=0.03 < \gamma_{\rm Y}$) is due to local plasticity inside the seed. 
The corresponding stress drop is at most of the order of the concentration of particles inside the seed, $c_{\rm I}$, which is less than 0.5 \%: see more details in Appendix~\ref{app:stress}.
Locations with a large nonaffine displacement, $D_{\rm min}^2$~\cite{falk1998dynamics}, correspond to the core of an Eshelby quadrupolar displacement field~\cite{tanguy2006plastic} taking place much before global failure. For small seeds, {\it e.g.}, $D_a=16$, a single quadrupolar displacement is found, but as $D_a$ is increased, we see two (for $D_a=32$) or three (for $D_a=50$) quadrupolar relaxations aligned along the major axis of the seed.

\begin{figure}[t]
\includegraphics[width=0.49\columnwidth]{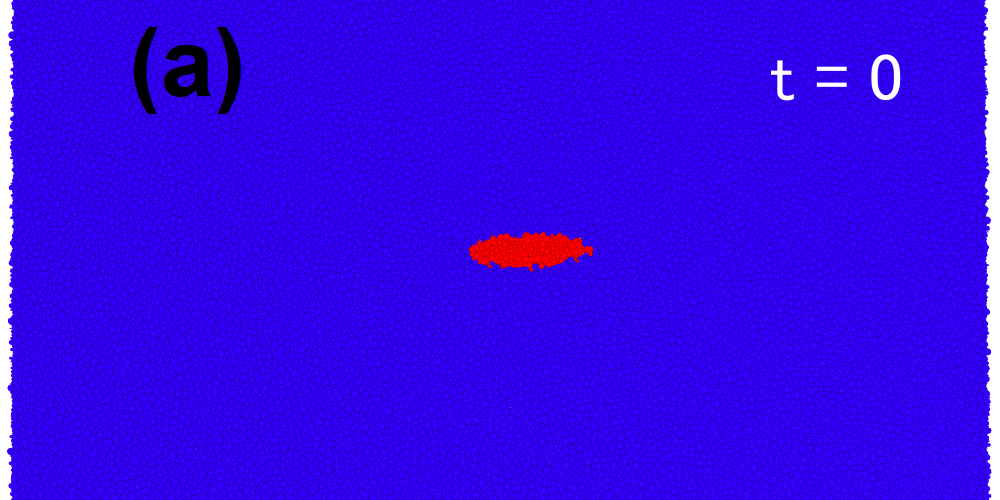}
\includegraphics[width=0.49\columnwidth]{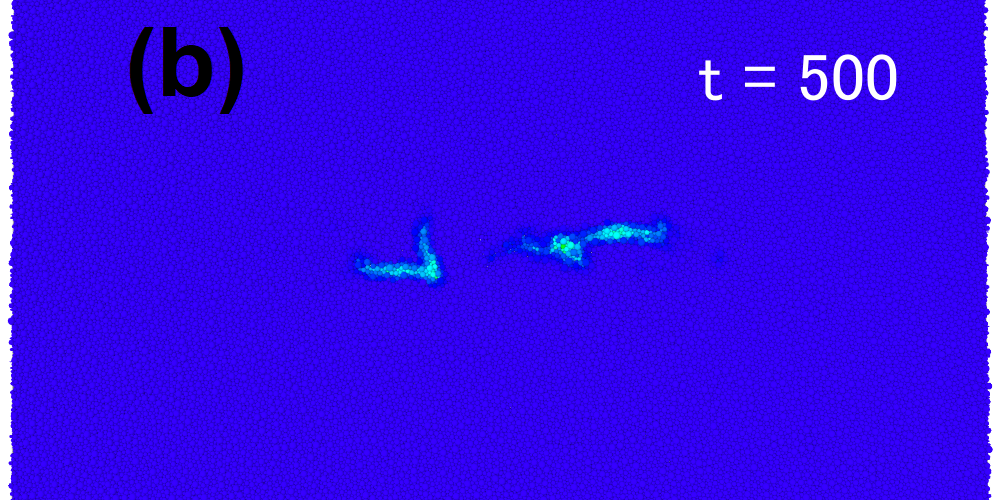}
\includegraphics[width=0.49\columnwidth]{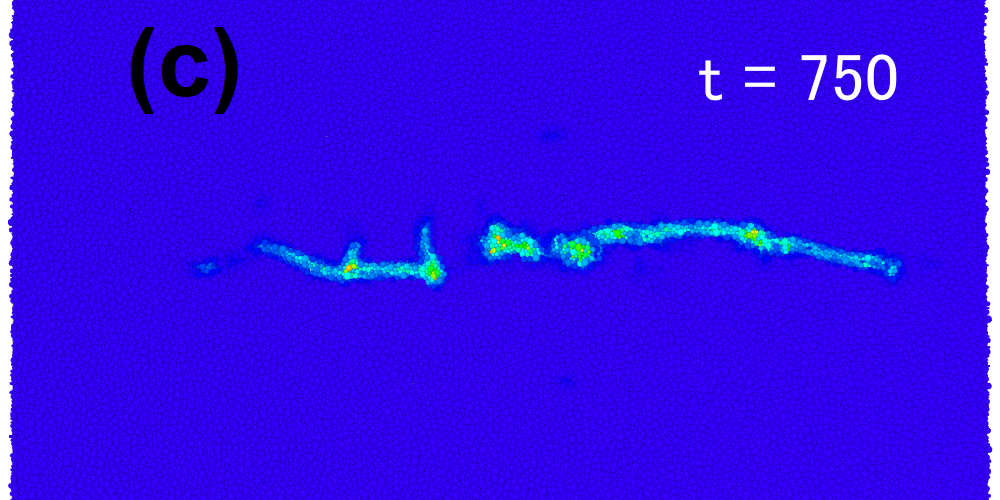}
\includegraphics[width=0.49\columnwidth]{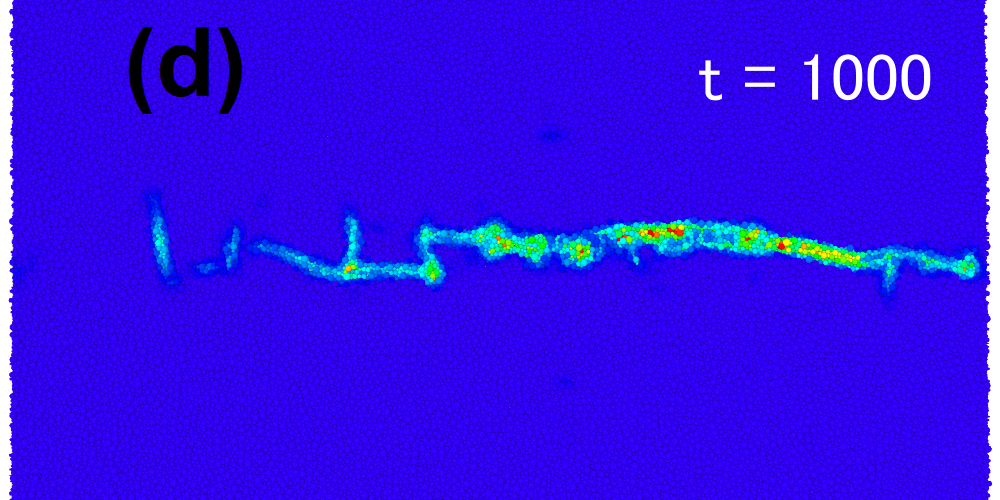}
\includegraphics[width=0.49\columnwidth]{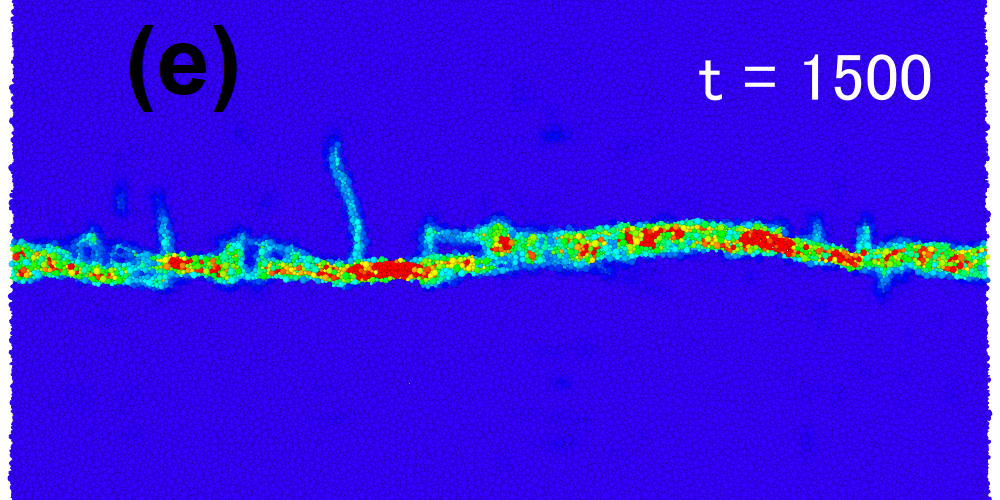}
\includegraphics[width=0.49\columnwidth]{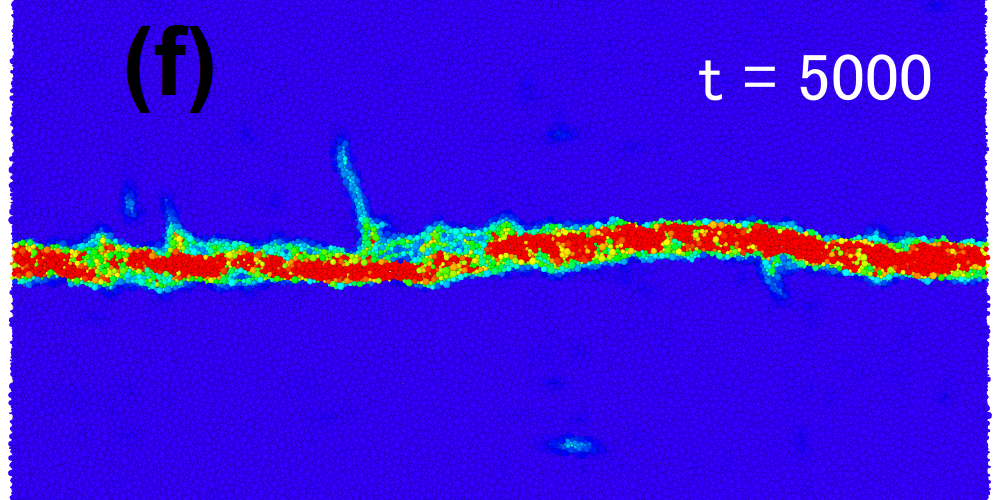}
\includegraphics[width=0.9\columnwidth]{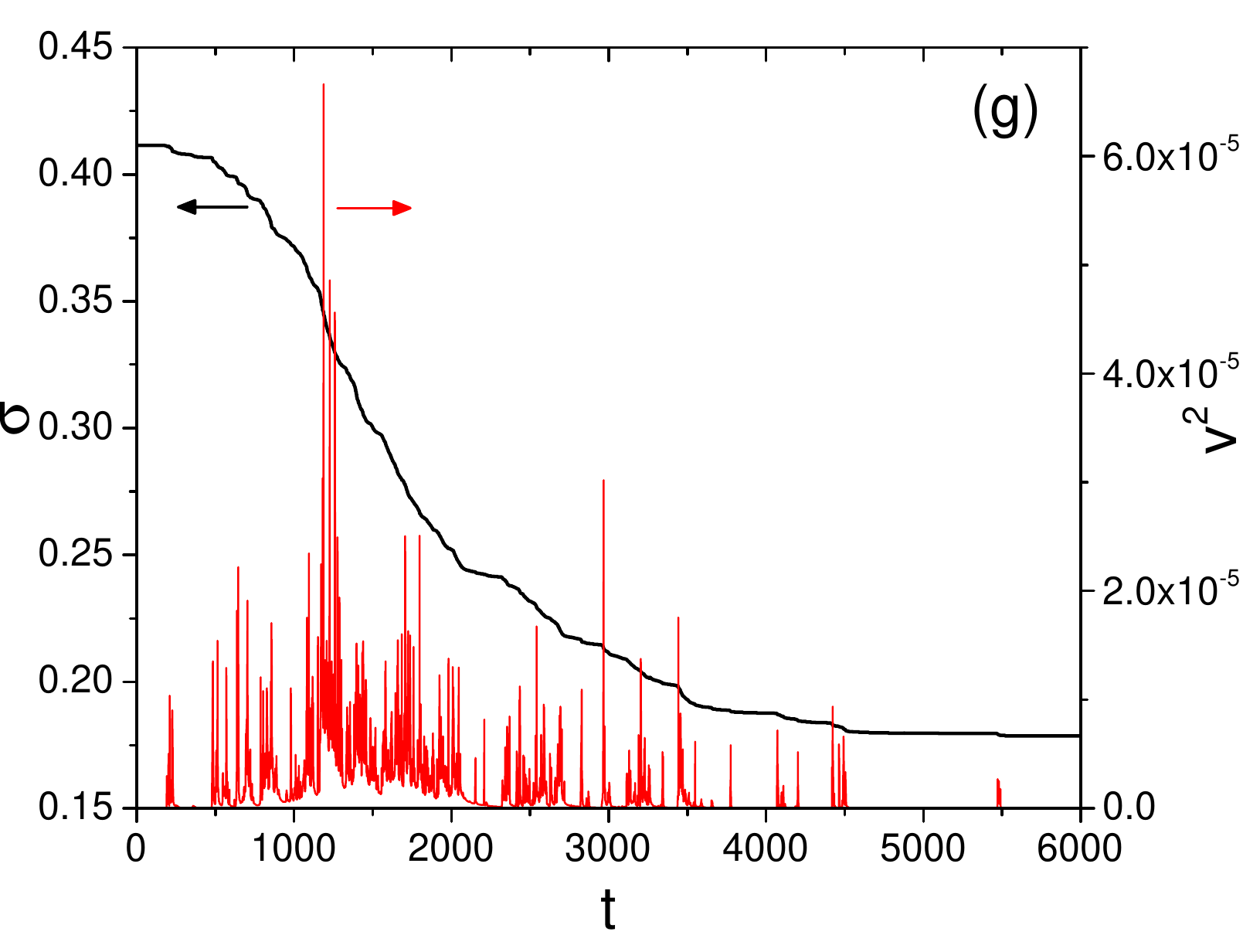}
\caption{Dynamics of shear-band formation for a stable glass with $T_{\rm ini}=0.035$ and $D_a=32$.
Top (a-f): Evolution of nonaffine square displacement $D_{\rm min}^2$ between $t=0$ and various times $t$ during the gradient-descent dynamics at yielding. 
At $t=0$, particles inside the seed are colored in red. 
Bottom (g): Corresponding intermittent time evolution of the stress $\sigma$ (left axis) and of the average squared velocity $v^2$ (right axis).}
\label{fig:during_yielding}
\end{figure}

The symmetry of the stress relaxation associated with these events implies that the redistributed stress is maximal at the tip of the seed, which eventually leads to the sudden formation of a macroscopic shear band. To analyze the physical process behind the ensuing stress drop of order $1$ associated with this brittle yielding, we perform gradient-descent dynamics starting from the configuration exactly on the verge of yielding. The time evolution of $D_{\rm min}^2$ during the stress drop is visualized in Fig.~\ref{fig:during_yielding}(a-f). Plastic events with large $D_{\rm min}^2$ first appear very near the tip of the seed, and then proliferate along the horizontal direction, ending up in a system-spanning shear band. For a uniform glass sample with no seed, the initial location of the shear band is random in the simulation box and its direction is either horizontal or vertical~\cite{kapteijns2019fast}. On the contrary, in the presence of a long enough seed (see Appendix~\ref{app:location} for a quantitative discussion and a comparison with spherical seeds), the location as well as the direction of the system-spanning shear band are fully determined by the seed. 

We also show the time evolution of the stress during the gradient descent in Fig.~\ref{fig:during_yielding}(g). Considering the relatively smooth decrease of $\sigma$, one might naively conclude that shear-band formation is merely the continuous sliding of two rigid blocks. To characterize more precisely the dynamical deformation process of the shear band formation, we have measured the squared velocity $v^2(t)=\frac{1}{N} \sum_{i} |{\bf v}_i(t)|^2$. This quantity is more sensitive to dynamical activity during the shear-band formation. The time dependence of $v^2(t)$ in Fig.~\ref{fig:during_yielding}(g) reveals that the deformation process is in fact highly intermittent. 
As shown in Appendix~\ref{app:propagation}, this intermittent behavior translates in real space into individual Eshelby-like displacements that repeatedly appear along the shear band. It is 
these multiple Eshelby relaxations that eventually lead to the stress drop of order 1. Since a single Eshelby event can only carry tiny displacements at its core, to have a 
macroscopic effect that is visible on the $\sigma$ vs $\gamma$ curve, Eshelby events have to appear repeatedly in the already formed shear band, as theoretically argued~\cite{dasgupta2013yield}, and confirmed in our simulations. 

\section{Discussions}

We can now rationalize our numerical findings by the following physical picture that is based on the idea that rare soft regions, and more specifically the largest of the softest ones appearing with a nonzero probability in the thermodynamic limit, play a crucial role in leading to shear banding for well-annealed glasses. By inserting by hand such regions, which in macroscopically large samples would be randomly located in space, we found that they rearrange much before the rest of the material. This local plastic 
activity leads to the formation and alignment of Eshelby-like quadrupoles inside the seeds. (Plasticity {\it inside} the soft seed thus plays a  crucial role, at variance with what happens for a crack or an inclusion.) The key point to then explain shear banding is that the aligned Eshelby quadrupoles induce contributions to the local stress that add up constructively along the direction of the seed's major axis, and only there. Due to the decay of the elastic interaction with distance (as $1/r^d$), the local stress generated by Eshelby relaxations is maximum near the tips of the ellipsoidal seed. We have confirmed numerically (see Appendix~\ref{app:local}) that the stress increase near the tips is of order $1$ during the initial shear-band 
propagation. This positive interference is at the root of shear banding. Indeed, provided the system is relatively uniform outside the seed, 
which should be the case for well-annealed glasses in which the disorder strength is small, the tips should thus yield first upon further straining and then induce 
an even larger stress near the new tips. A self-sustained process and a system-spanning shear band naturally ensues. 

The above picture corresponds to a ``weak-disorder'' regime in which fluctuations of the local yield stress are relatively small. Disorder is nonetheless crucial 
because it allows the spontaneous, albeit extremely rare, presence of a large soft region. Disorder also gives rise to the intermittent behavior during shear 
banding shown in Fig.~\ref{fig:during_yielding}. For larger disorder strength (less stable glasses) the shear-band propagation should take more complex paths due to the stronger nonuniformity of the material: strong pinning sites where the local yield stress may be high must be avoided while the presence of 
additional soft spots~\cite{richard2020predicting} may be used. The propagating shear band may then need to deform substantially to go through the softest regions which are no longer exactly at its tips~\cite{ozawa2020role}. 
What happens for still larger disorder is unclear and will be the focus of a subsequent work. It has been argued that the sharp stress drop cannot be replaced by an overshoot in this regime, but only by a monotonously increasing behavior. The reason is that a long-wavelength linear instability is generated whenever the stress-strain curve displays a continuous overshoot and could then lead to a brittle yielding~\cite{moorcroft2013criteria,barlowductile,fielding2021yielding}. (For well-annealed samples, this linear instability 
is preempted by the discontinuity triggered by the rare events studied here; in fact the discontinuous stress jump appears for a strictly positive slope of the stress-strain curve). This argument however does not take into account that disorder can pin the propagation of the instability and hence preserve a 
continuous overshoot behavior question.
This open question deserves further numerical and analytical studies~\cite{richard2021finite}. 

We now go back to the strong finite-size effects of the stress versus strain curve for well-annealed, brittle, samples and the proper thermodynamic limit. The 
limiting value of the shear stress at the discontinuous yielding, $\sigma_{\rm Y}$, is determined by the stress needed to propagate a shear band from 
an arbitrary large seed embedded in a much larger, macroscopic, sample. To realize this scenario numerically, we should in principle analyze the double limit of 
$1 \ll D_a \ll L$. This explains why we have limited the seed length $D_a$ so that the tips of the seed are not affected by their images in the periodically repeated simulation box. In practice, we have chosen the longest seed to be $D_a=50$, which is about half the smallest system size $L$ considered ($N=8000$ and $L=89.4$), and we have studied the variation at {\it fixed} $D_a$ when increasing $L$: more details are given in Appendix~\ref{app:finite}. For a large seed immersed in an even larger piece of material we expect that $\sigma_{\rm Y}$ does not correspond to the stress in the steady-state regime obtained at large $\gamma$. The latter is the stress 
needed to induce plastic activity after the shear band has already formed and propagated through the entire system. A larger external stress instead is needed for the stress at the tip to reach its local yield stress value, as this still 
represents the yield stress of a yet unrelaxed stable glass region. We conclude that a nonzero stress drop of order $1$ at yielding survives in the thermodynamic limit. A different conclusion, based on fracture mechanics, was reached in Ref.~\cite{popovic2018elastoplastic}~\footnote{Note that the derivation from fracture mechanics predicting stress and strain in the presence of a crack requires that linear elasticity holds. This is however not true for a shear-band seed that yields prior to the bulk and thereby generates plastic components to the stress and strain fields.}.

As already emphasized, introducing large soft regions by hand is a mean to get around the strong finite-size effects associated with their spontaneous occurrence. The 
probability of such a region of volume $v$ anywhere in a system of volume $L^d$ goes as $(L^d/v)\exp[-(c/\xi^d)v]$ and is of order 1 only for $L$ exponentially large in $v$.  
As then illustrated in Fig.~\ref{fig:stress}, this can explain why the yield strain $\gamma_{\rm Y}$ increases with decreasing system size in deformed sub-micron metallic-glass samples~\cite{yang2012size,tian2012approaching}. In turn, we have also shown that seed influence fades away for poorly annealed glasses. Finite-size effects 
resulting from the potential presence of a long-wavelength instability~\cite{barlowductile,fielding2021yielding,richard2021finite} must then be addressed by other means. 
To make progress on these fundamental issues, the main challenge is to develop a theory of the 
propagation and growth of shear bands, whether triggered by a rare soft region or by a long-wavelength instability, within an amorphous solid. The phenomenon presents 
analogies with the depinning of interfaces in a random medium, yet with some crucial differences due to the anisotropic and long-range character of the elastic interactions 
and to the very nature and shape of the propagating object. 

\begin{acknowledgments}
We thank S. Fielding and M. Wyart for discussions. This work was supported by grants from the Simons Foundation (\#454933 Ludovic Berthier and \#454935 Giulio Biroli).
\end{acknowledgments}

\appendix

\section{Details on the simulation methods}

\label{sec:method_SI}

{\it Model}. 
The two-dimensional glass-forming model consists of particles with purely repulsive interactions and a continuous size polydispersity. Particle diameters, $d_i$, are 
randomly drawn from a distribution of the form: $f(d) = Ad^{-3}$, for $d \in [ d_{\rm min}, d_{\rm max} ]$, where $A$ is a normalization constant. The size polydispersity 
is quantified by $\delta=(\overline{d^2} - \overline{d}^2)^{1/2}/\overline{d}$, where the overline denotes an average over the distribution $f(d)$. Here we choose 
$\delta = 0.23$ by imposing $d_{\rm min} / d_{\rm max} = 0.449$. The average diameter, $\overline{d}$, sets the unit of length. The soft-disk interactions are pairwise 
additive and described by an inverse power-law potential
\begin{eqnarray}
v_{ij}(r) &=& v_0 \left( \frac{d_{ij}}{r} \right)^{12} + c_0 + c_1 \left( \frac{r}{d_{ij}} \right)^2 + c_2 \left( \frac{r}{d_{ij}} \right)^4, \nonumber \\
d_{ij} &=& \frac{(d_i + d_j)}{2} (1-\epsilon |d_i - d_j|), \nonumber
\end{eqnarray}
where $v_0$ sets the unit of energy (and of temperature with the Boltzmann constant $k_\mathrm{B}\equiv 1$) and $\epsilon=0.2$  quantifies the degree of 
nonadditivity of particle diameters. We introduce $\epsilon>0$ in the model to suppress fractionation and thus enhance the glass-forming ability. The constants $c_0$, $c_1$ 
and $c_2$ enforce a vanishing potential and continuity of its first- and second-order derivatives at the cut-off distance $r_{\rm cut}=1.25 d_{ij}$.
We simulate a system with $N$ particles within a square cell of area $V=L^2$ where $L$ is the linear box length, under periodic boundary conditions, at a number density 
$\rho=N/V=1$. We study $N$ from $8000$ to $64000$.

We compute the shear stress $\sigma$ through 
\begin{equation}
    \sigma = \frac{1}{N} \sum_i \sum_{\substack{j \\ (j>i)}} \frac{x_{ij}y_{ij}}{r_{ij}} v_{ij}'(r_{ij}),
    \label{eq:shear_stress}
\end{equation}
where $v_{ij}'$ is the derivative of the potential.

\vspace{0.5cm}
{\it Preparation of samples with a seed}. 
Glass samples with a seed (soft region) have been prepared by first equilibrating liquid configurations at a finite temperature, $T_{\rm ini}$, which is sometimes referred to 
as the fictive temperature of the glass sample.  We prepare equilibrium configurations for the polydisperse spheres using swap Monte-Carlo 
simulations~\cite{ninarello2017models}, which allows us to access a wide range of $T_{\rm ini}$. With probability $P_{\rm swap}=0.2$, we perform a swap move where 
we pick two particles at random and attempt to exchange their diameters, and with probability $1-P_{\rm swap}=0.8$, we perform conventional Monte-Carlo 
translational moves. 

In order to introduce a seed (soft region), we define an ellipsoidal region whose size is characterized by the length of the major axis 
 $D_a$ and the length of the minor axis $D_b$ (see Fig.~2 in the main text). For most of our study $D_b$ is fixed to the value of $8$, and we vary $D_a$ from $0$ to $90$. 
 The linear box length of the two-dimensional system for $N=8000$ is $L = 89.4$ and for $N=64000$ is $L = 253$.
We then perform additional swap Monte-Carlo simulations only for the particles inside the ellipsoidal region defined above while the particles outside stay pinned.
The temperature of this additional Monte-Carlo simulations is $T_{\rm h}=10.0$. The dynamical mode-coupling crossover~\cite{gotze2008complex} of the  system is 
$T_{\rm mct} \approx 0.110$. Thus $T_{\rm h}$ is about $100$ times higher than the mode-coupling crossover temperature of the system.
We then quench the obtained configuration down to zero temperature by using the conjugate gradient method. Thus our glass samples contain a disordered, 
poorly-annealed region in the middle of the simulation box (see Fig.~\ref{fig:before_yielding} in the main text).

\vspace{0.5cm}
{\it Mechanical loading}. 
We have performed strain-controlled athermal quasi-static shear (AQS) deformation using Lees-Edwards boundary conditions~\cite{maloney2006amorphous}.
The AQS shear method consists of a succession of tiny uniform shear deformation with $\Delta \gamma=10^{-4}$, followed by energy minimization via the conjugate-gradient 
method. The AQS deformation is performed along the $x$-direction. Note that during the AQS deformation, the system is always located in a potential energy minimum 
(except of course during the transient conjugate-gradient minimization), i.e., it stays at $T=0$.

\vspace{0.5cm}
{\it Nonaffine displacement}.
We consider the local nonaffine displacement of a given particle relative to its nearest neighbor particles, $D_{\rm min}^2$~\cite{falk1998dynamics}. $D_{\rm min}^2(\gamma_0,\gamma)$ is measured between the configuration at $\gamma_0$ and $\gamma$. We define nearest neighbors by using the cut-off radius of the interaction 
range, $R_{\rm cut}=3.0 \overline{d}$. We determine the nearest neighbors of a particle from the configuration at $\gamma_0$. For the gradient-descent dynamics, we 
replace $\gamma_0$ and $\gamma$ by two times $t_0$ and $t$ and consider $D_{\rm min}^2(t_0,t)$.

\section{Determination of $\gamma_{\rm Y}$}
\label{app:gammay}

\begin{figure}[htbp]
\includegraphics[width=0.48\columnwidth]{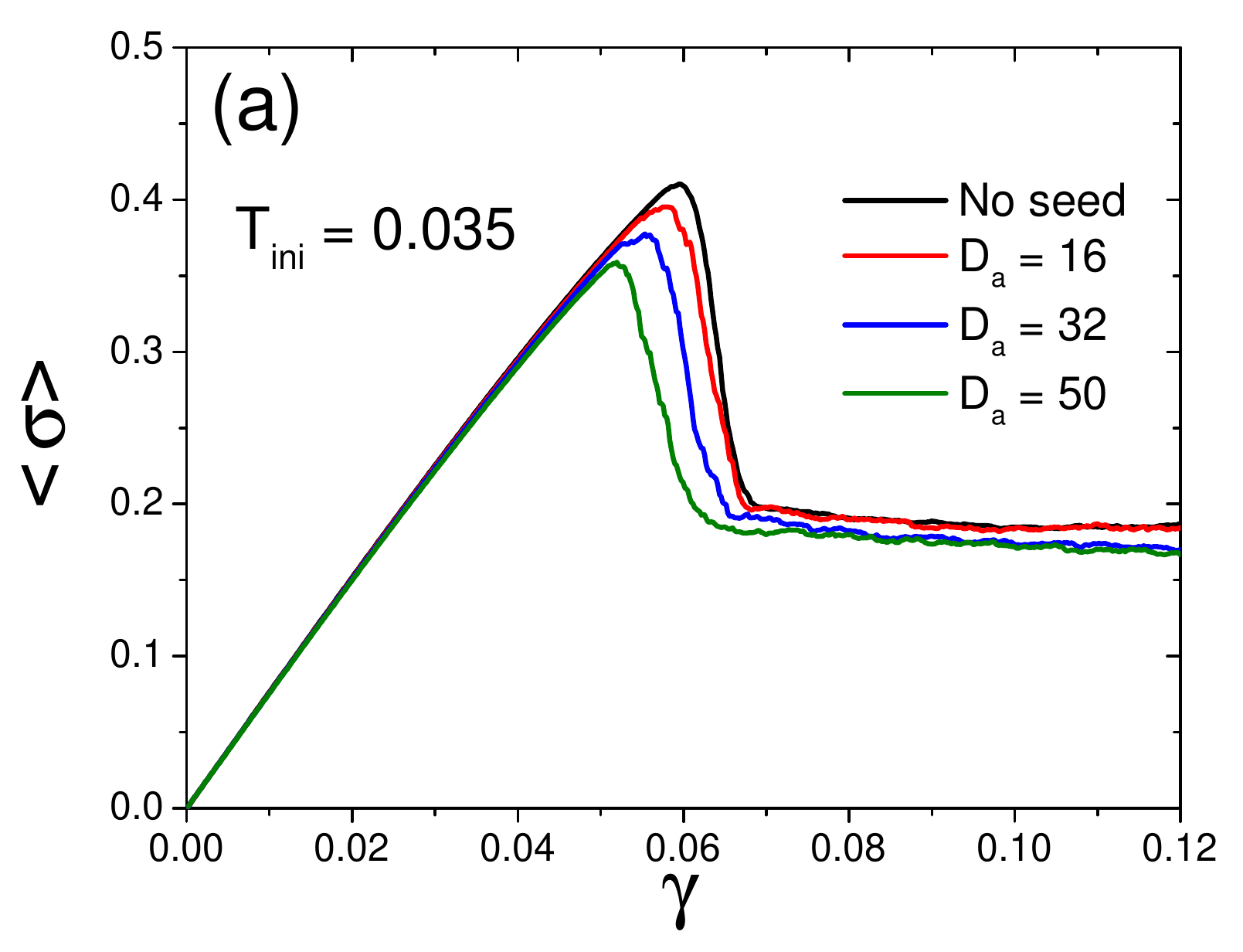}
\includegraphics[width=0.48\columnwidth]{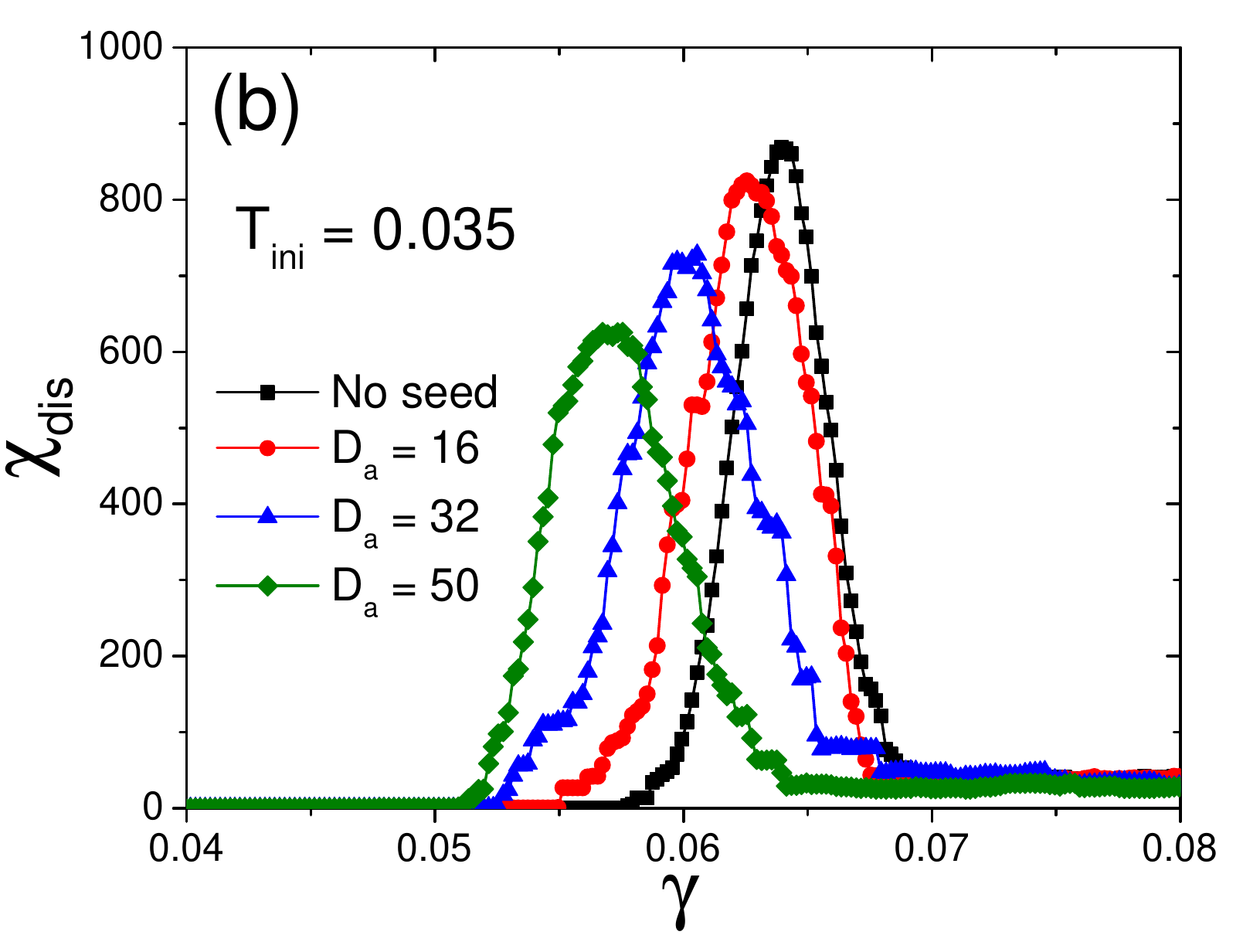}
\includegraphics[width=0.48\columnwidth]{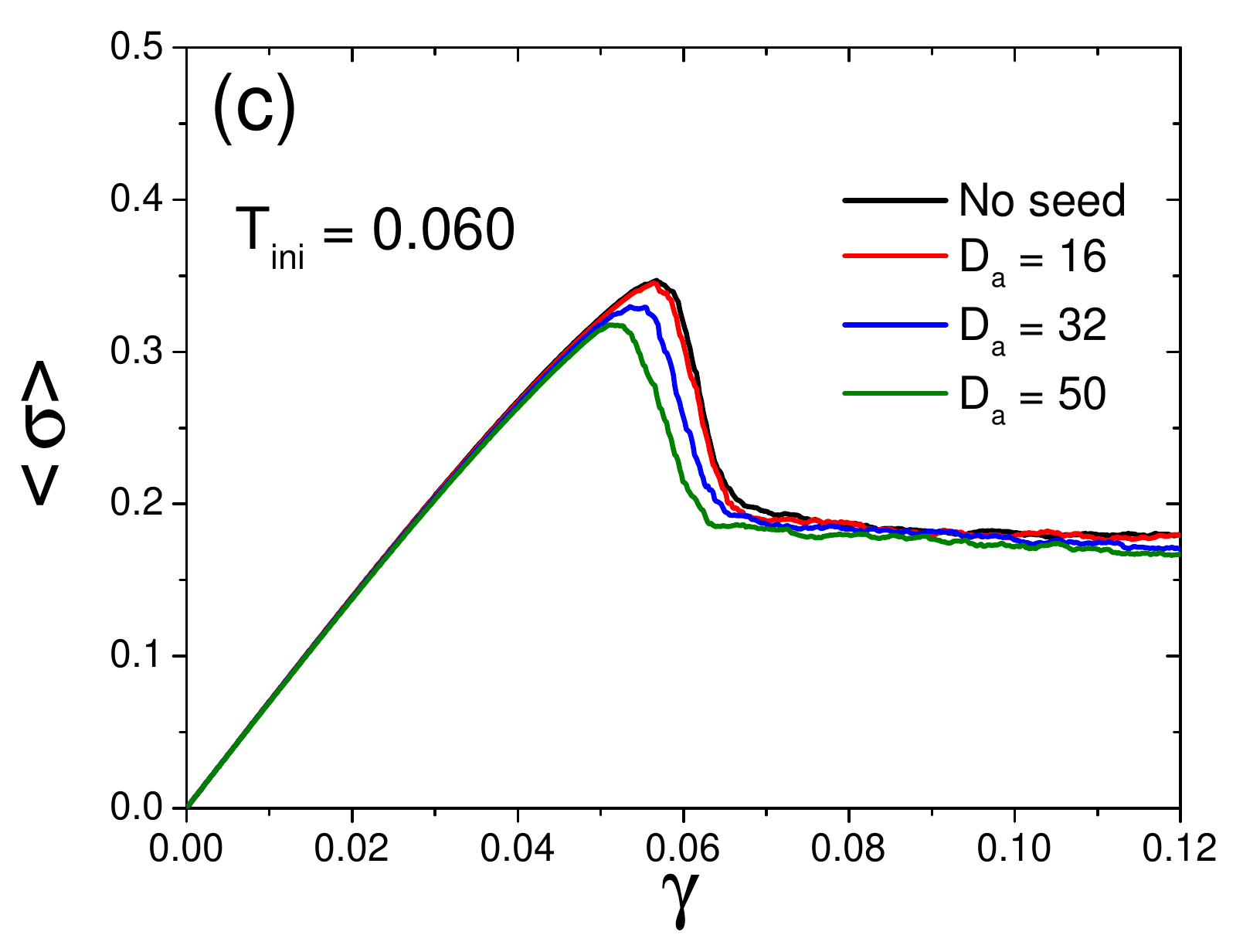}
\includegraphics[width=0.48\columnwidth]{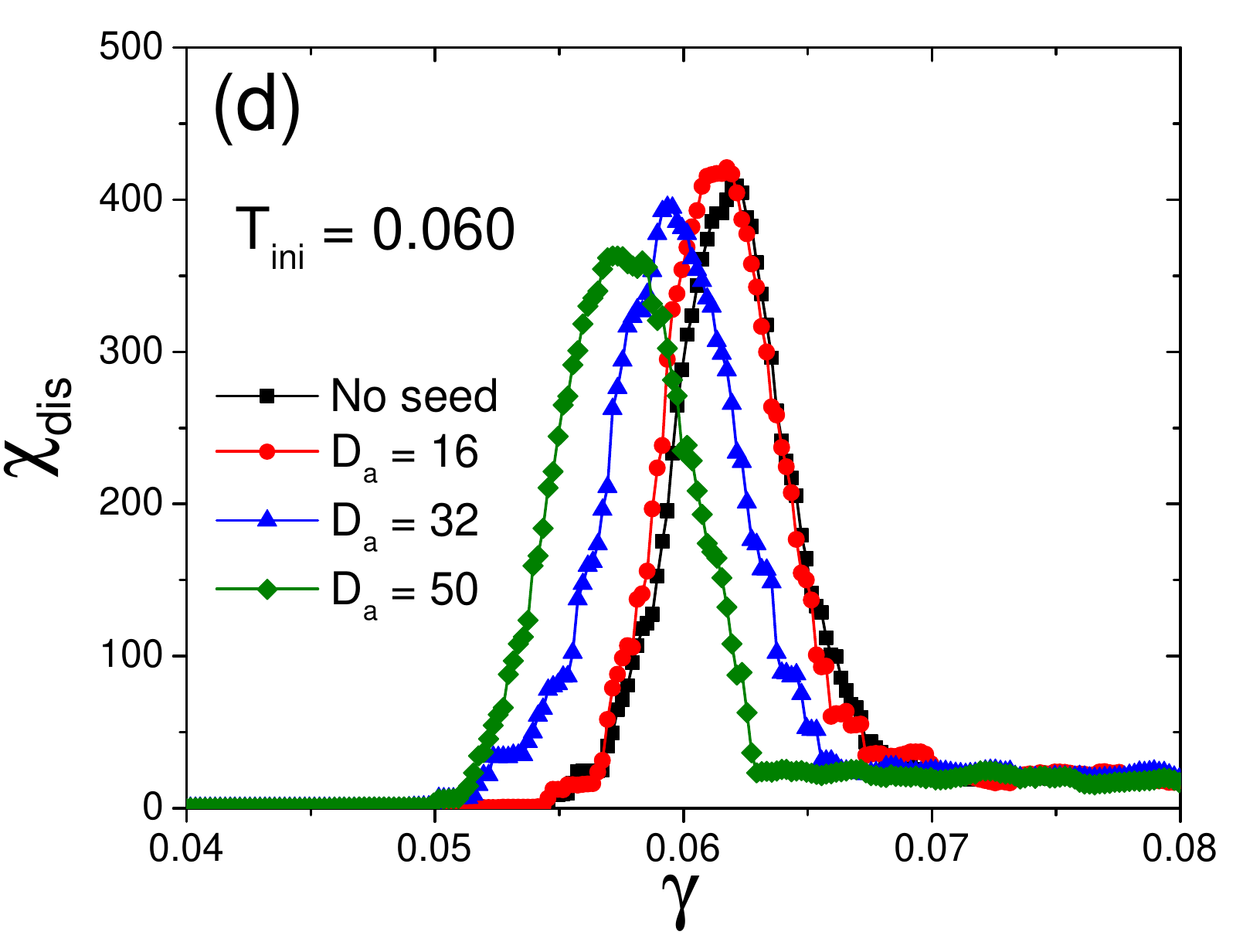}
\includegraphics[width=0.48\columnwidth]{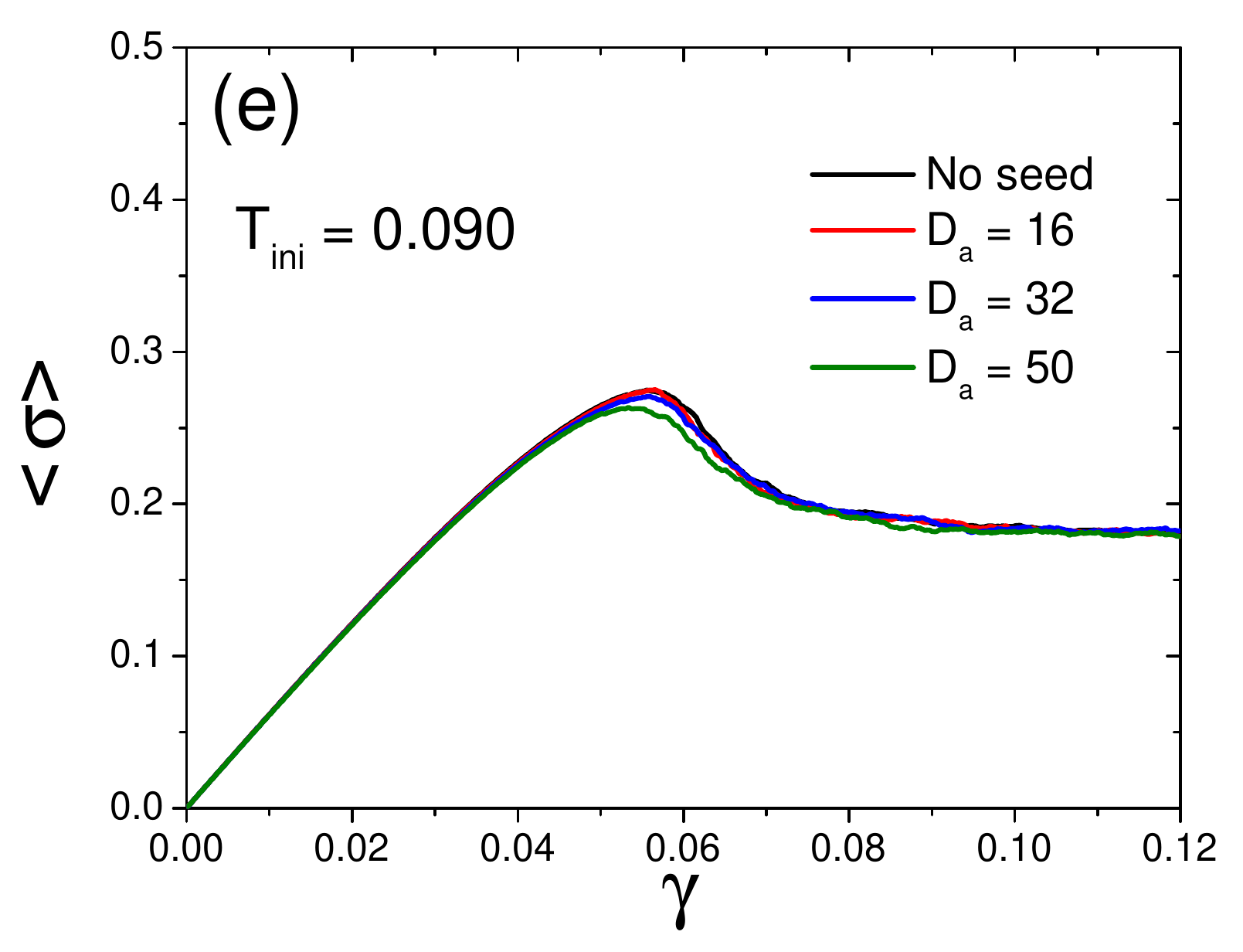}
\includegraphics[width=0.48\columnwidth]{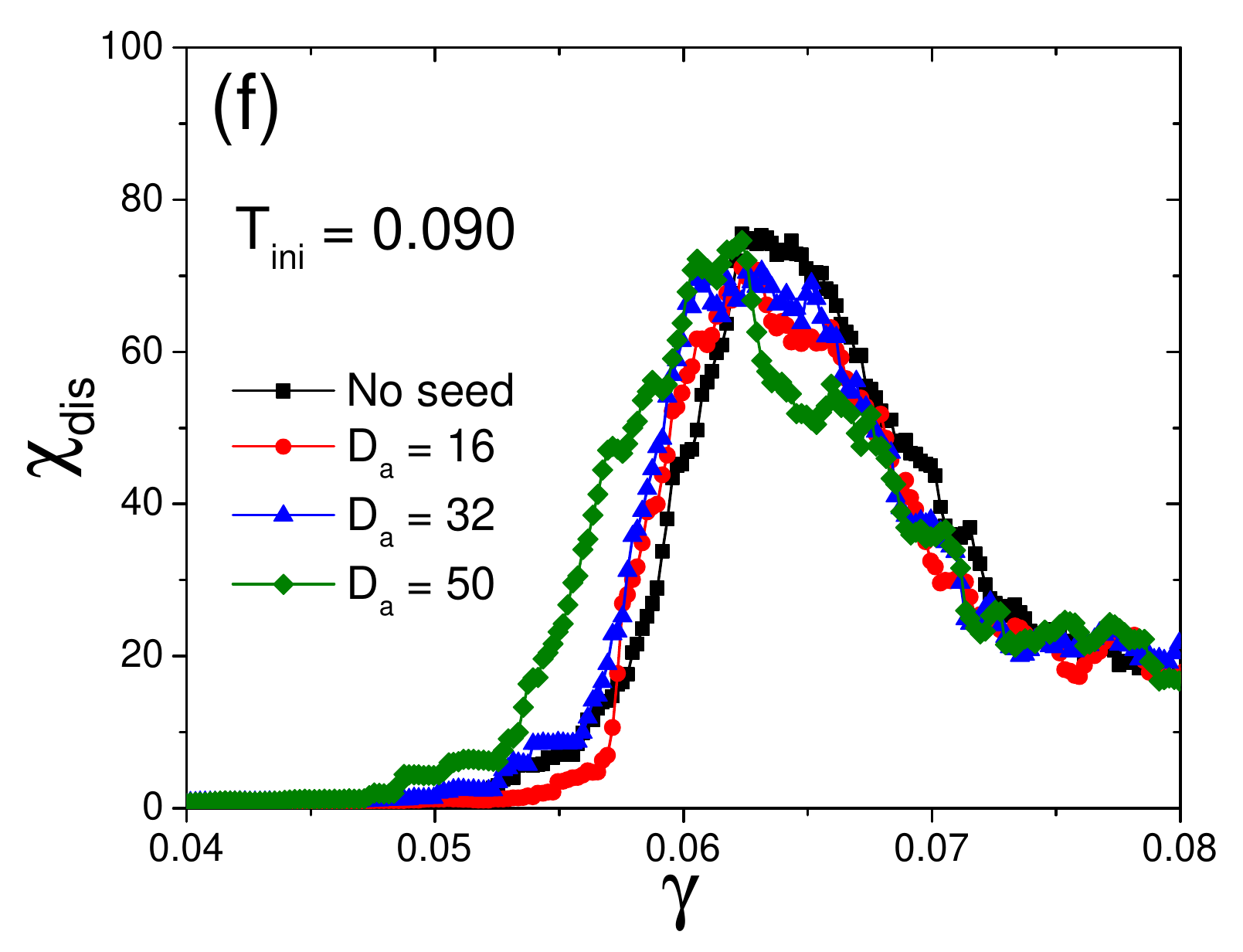}
\includegraphics[width=0.48\columnwidth]{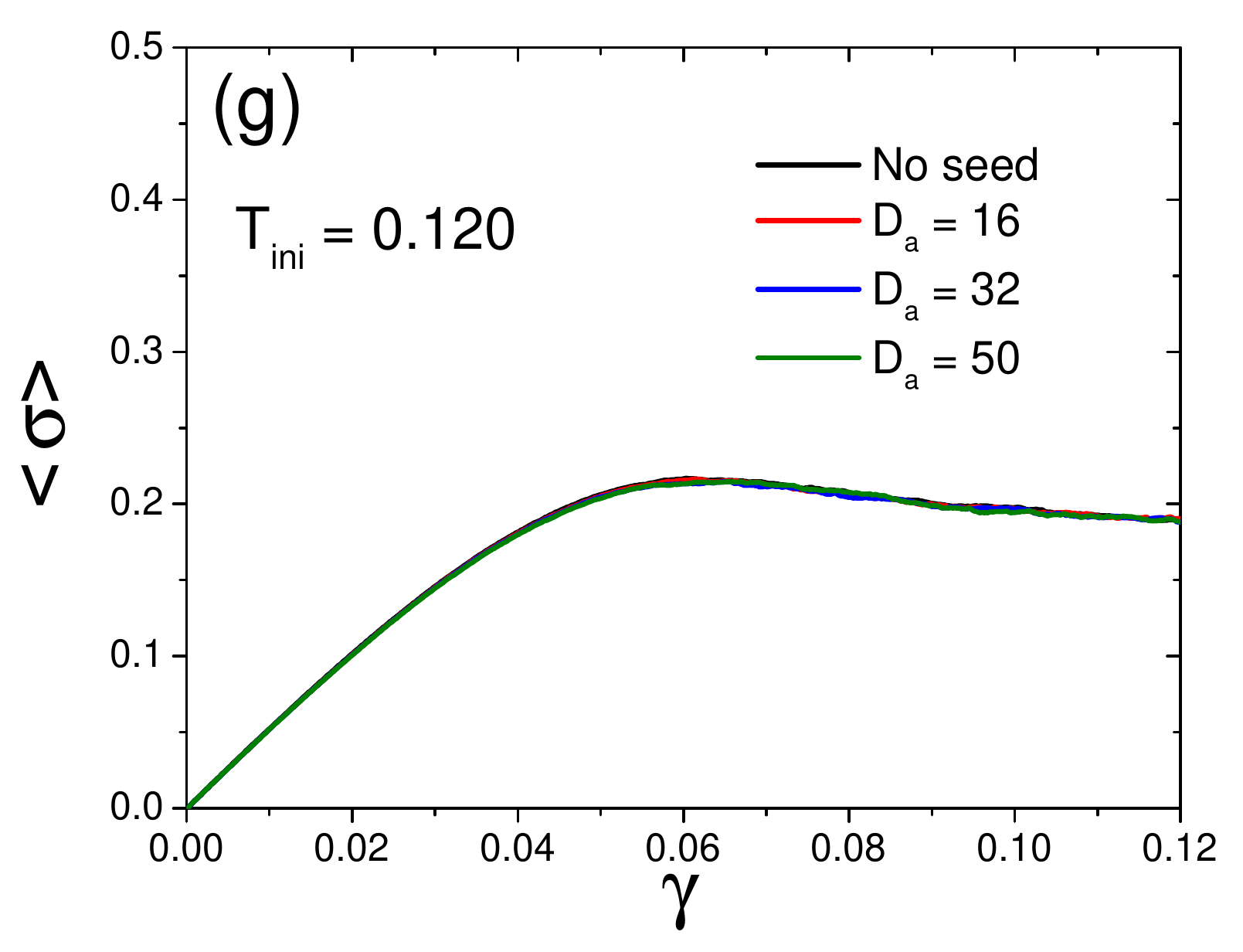}
\includegraphics[width=0.48\columnwidth]{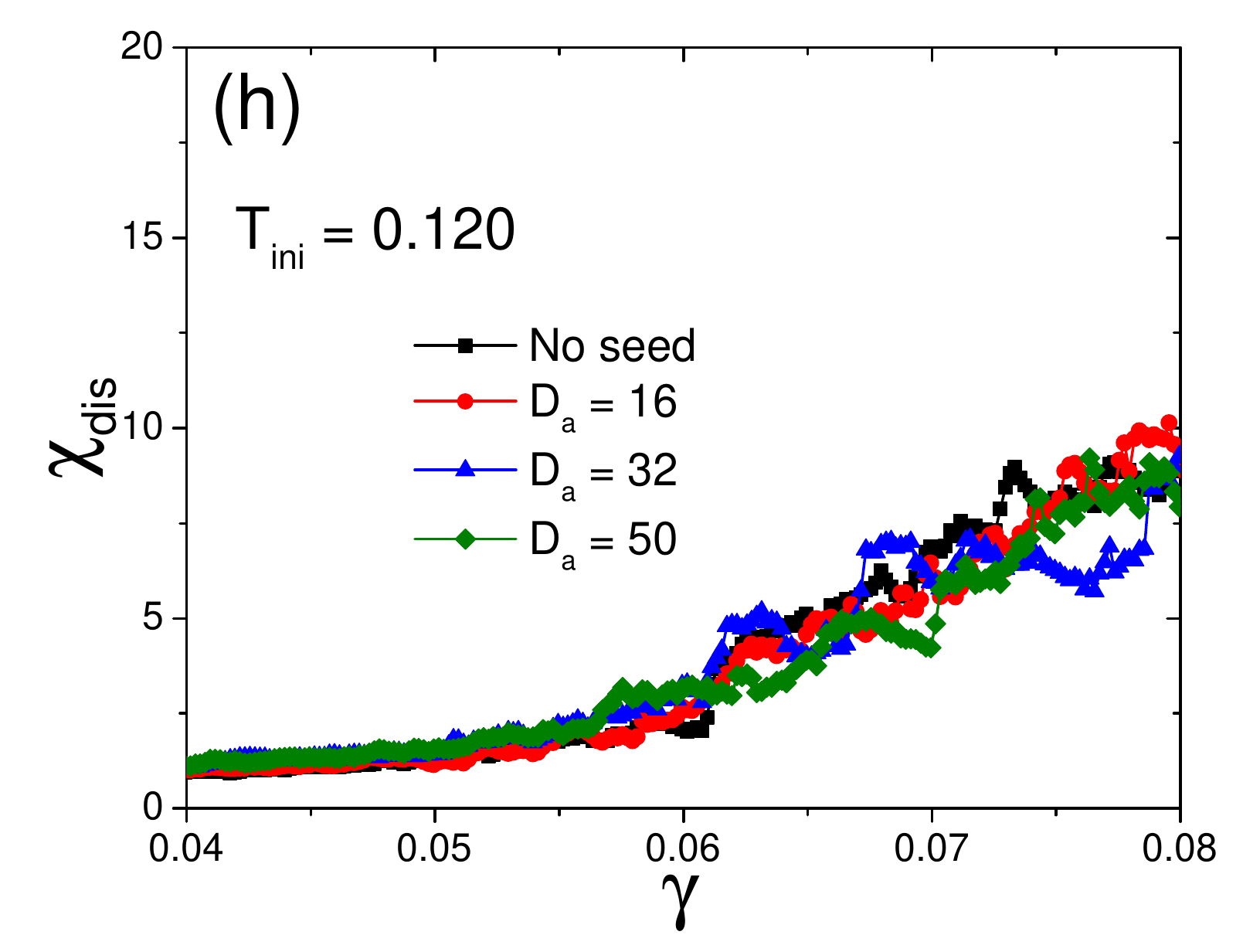}
\caption{The averaged stress $\langle \sigma \rangle$ (a,c,e,g) and the corresponding variance, $\chi_{\rm dis}=N(\langle \sigma^2 \rangle - \langle \sigma \rangle^2)$ 
(b,d,f,h), in the two-dimensional $N=64000$ systems for several $T_{\rm ini} $'s and $D_a$'s.
}
\label{fig:disconnected_2D}
\end{figure}

We explain how to determine the yield strain $\gamma_{\rm Y}$ for each $T_{\rm ini}$ and $D_a$.
In Fig.~\ref{fig:disconnected_2D}, we show the mean stress $\langle \sigma \rangle$ and the corresponding variance, 
$\chi_{\rm dis}=N(\langle \sigma^2 \rangle - \langle \sigma \rangle^2)$, averaged over $100-400$ independent samples. For stable glasses, say $T_{\rm ini}=0.035$, the 
macroscopic stress drop takes place earlier as one increases $D_a$. Thus, the remarkable phenomenology shown in Fig.~\ref{fig:stress} in the main text for individual samples 
is also confirmed at the ensemble level. The corresponding variance, $\chi_{\rm dis}$, shows a noticeable peak near the location of the macroscopic stress drop, 
and it shifts with increasing $D_a$. We can therefore use the peak position of $\chi_{\rm dis}$ as an unambiguous way to determine the yield strain, $\gamma_{\rm Y}$.
As $T_{\rm ini}$ is increased ($T_{\rm ini}=0.060$), the shift in both $\langle \sigma \rangle$ and $\chi_{\rm dis}$ plots becomes weaker. When $T_{\rm ini}$ is 
further increased ($T_{\rm ini}=0.090-0.120$), the $D_a$-dependence is significantly suppressed. Above a value near the apparent critical point, 
$T_{\rm ini, c}=0.085$, of the $N=64000$ system~\cite{ozawa2020role}, the $D_a$-dependence essentially disappears, which means that the seed does not play any 
role in the ductile yielding regime. 

\section{Location and direction of the shear band}

\label{app:location}

\begin{figure}[htbp]
\includegraphics[width=0.48\columnwidth]{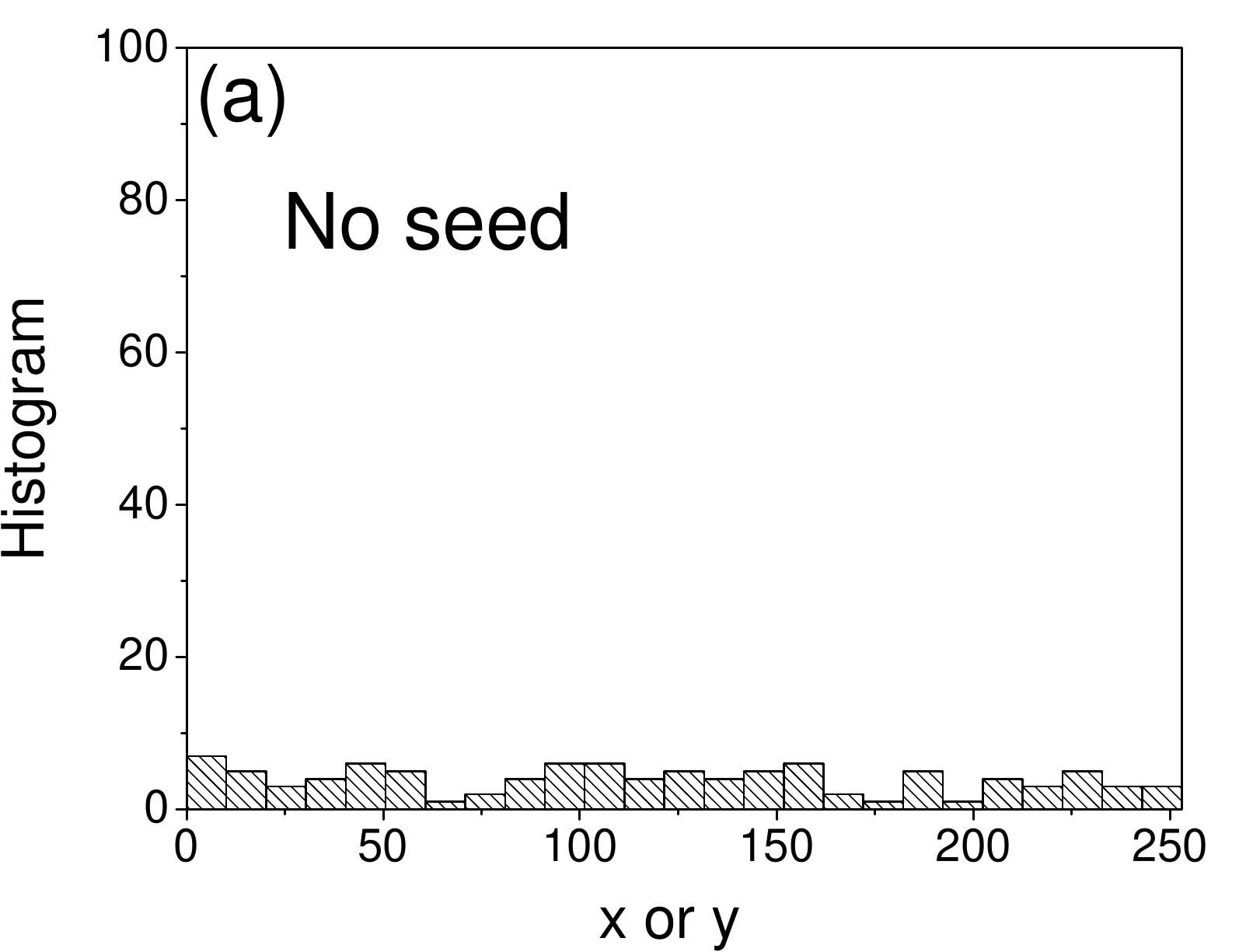}
\includegraphics[width=0.48\columnwidth]{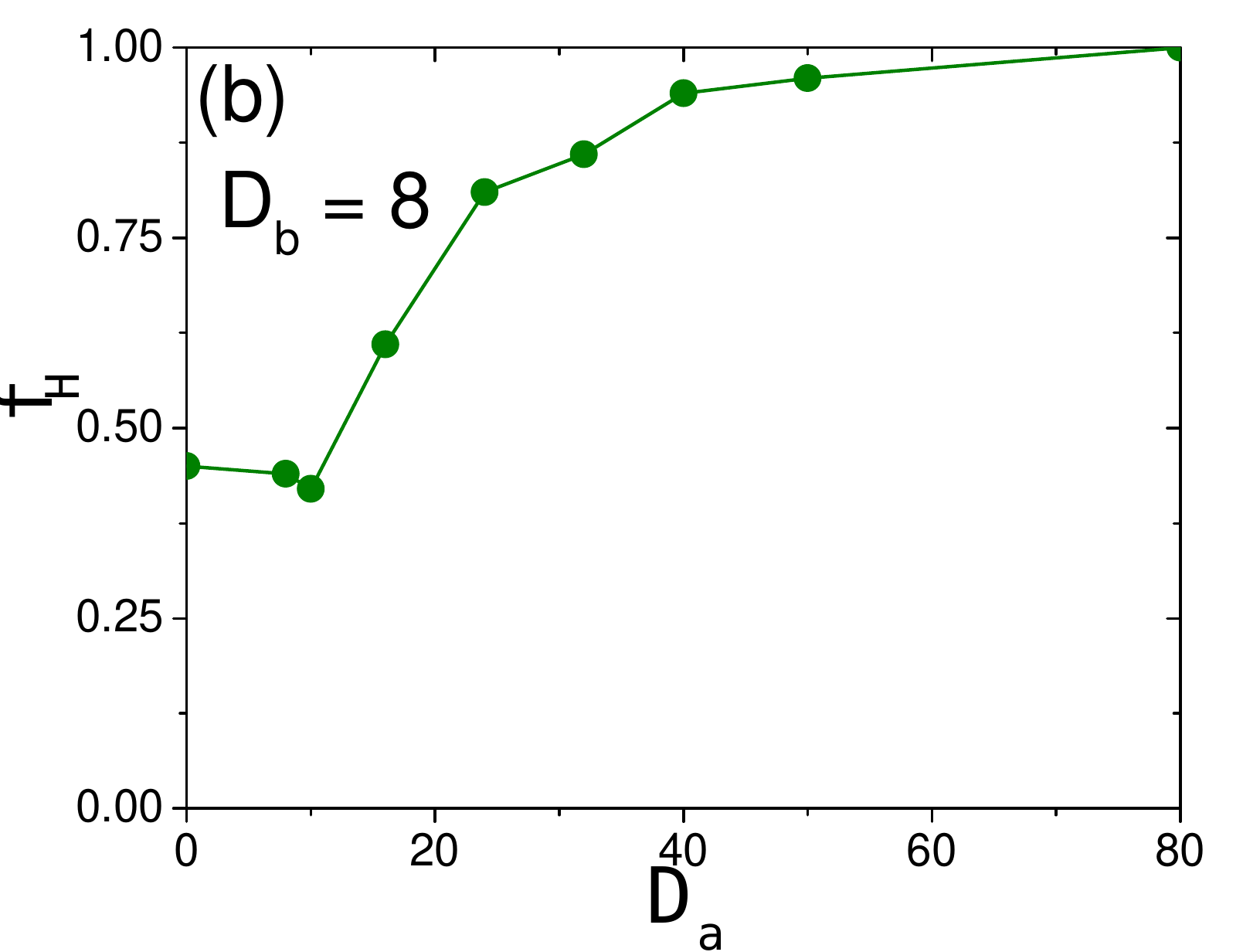}
\includegraphics[width=0.48\columnwidth]{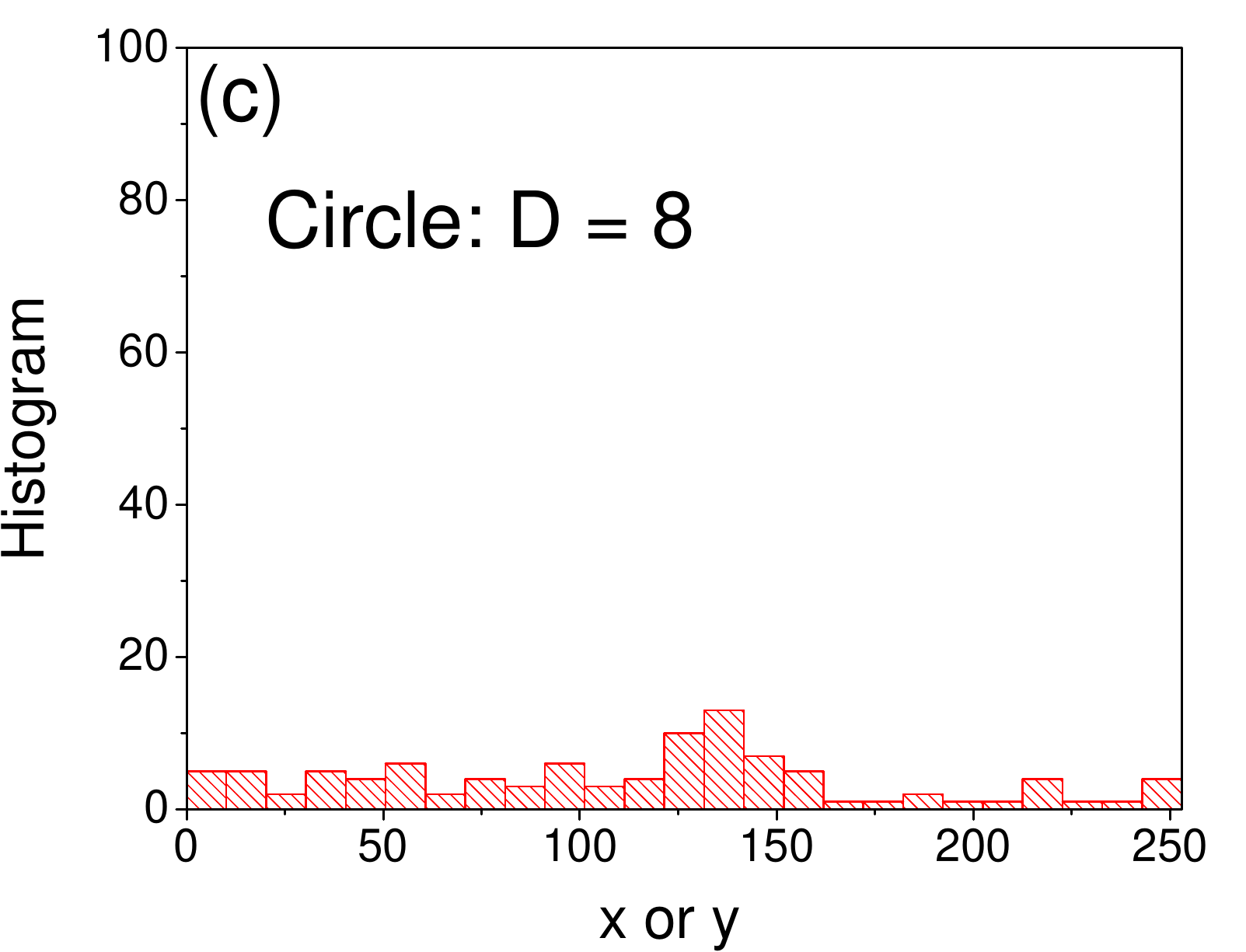}
\includegraphics[width=0.48\columnwidth]{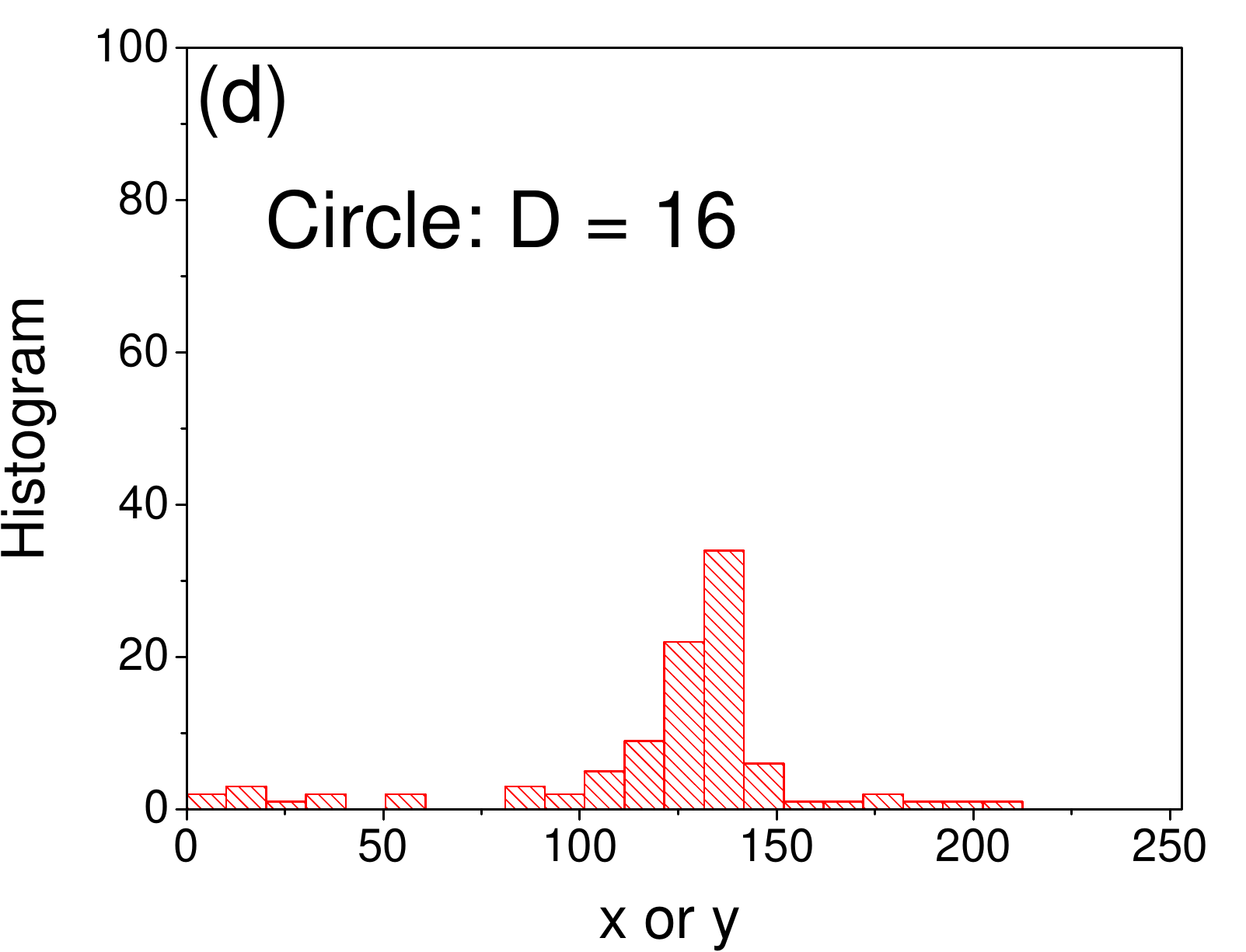}
\includegraphics[width=0.48\columnwidth]{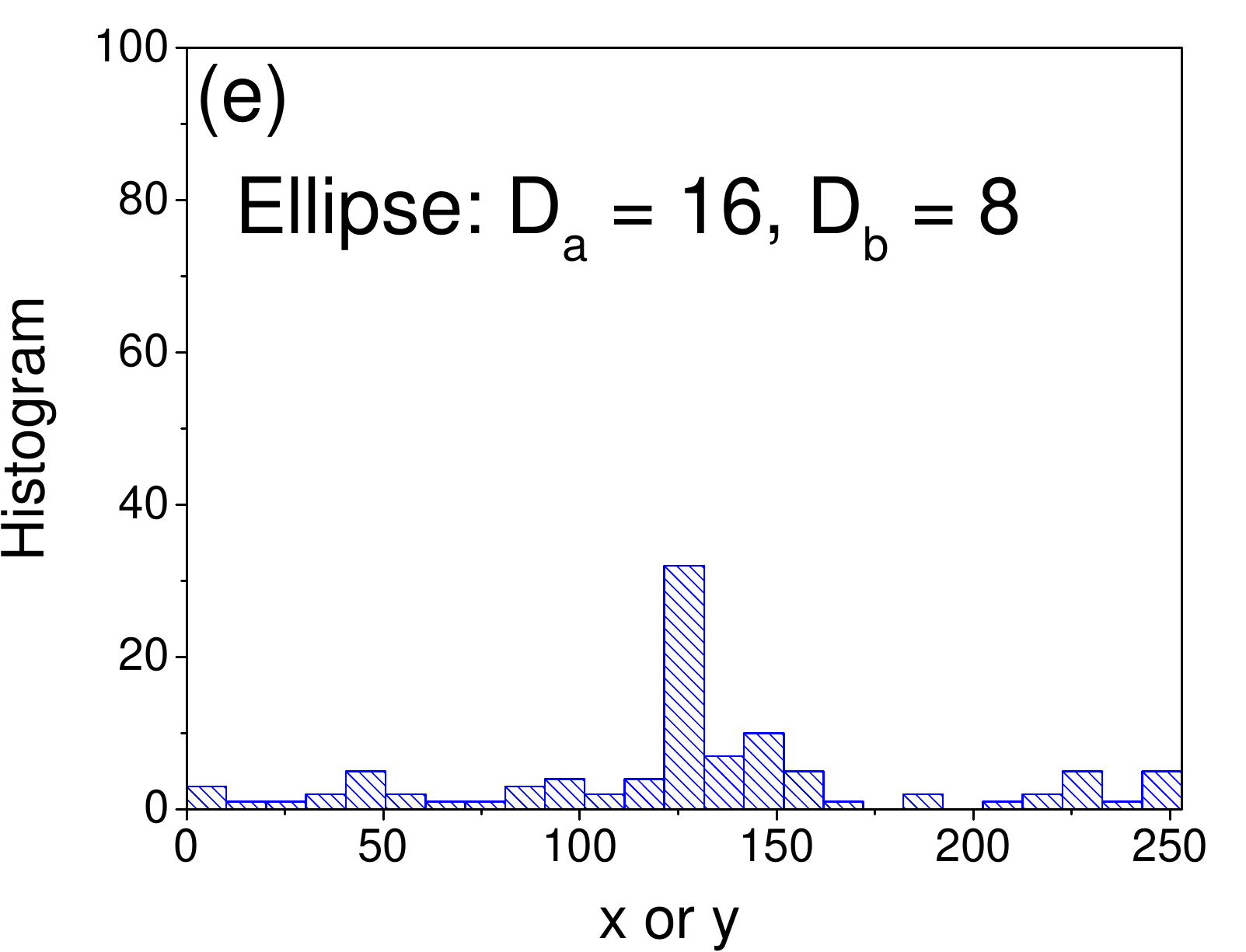}
\includegraphics[width=0.48\columnwidth]{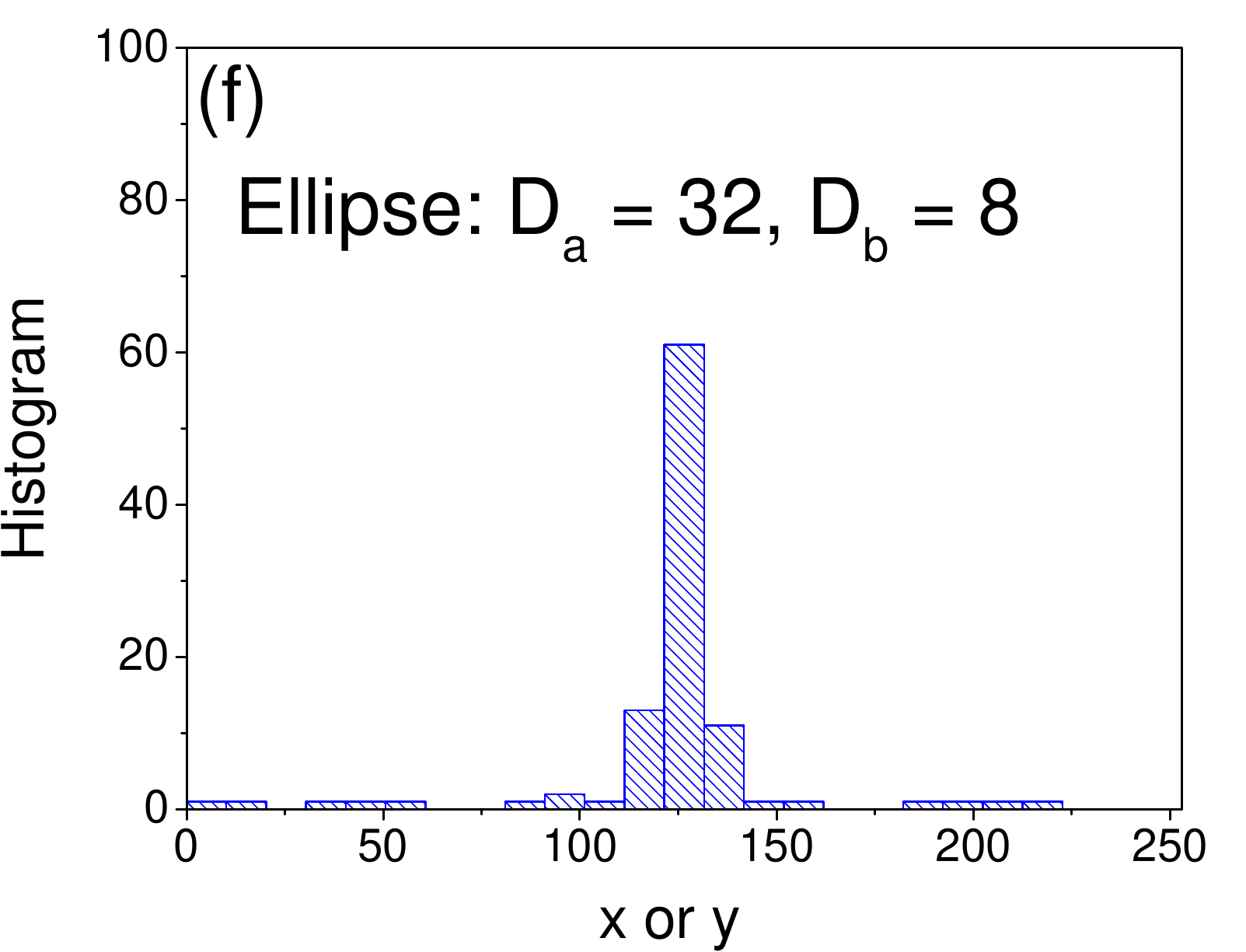}
\caption{
(a) Histogram of the location of the shear band. The fraction of samples with a horizontal shear band is $f_{\rm H}\approx0.45$.
(b) $f_{\rm H}$ as a function of $D_a$ with fixed $D_b=8$.
(c) Circular seed with $D_{\rm a}=D_{\rm b}=8$: $f_{\rm H}=0.44$. (d) Circular seed with $D_{\rm a}=D_{\rm b}=16$: $f_{\rm H}=0.52$. 
(e) Ellipse with $D_{\rm a}=16$ and $D_{\rm b}=8$: $f_{\rm H}=0.61$. (f) Ellipse with $D_{\rm a}=32$ and $D_{\rm b}=8$: $f_{\rm H}=0.86$.
}
\label{fig:histgram}
\end{figure}

Here we show that the presence of a soft ellipsoidal region in amorphous solids determines the location and direction of the shear band (hence the term ``seed'').
For stable homogeneous samples without a seed, the shear band occurs in any place under the Lees-Edwards periodic boundary conditions. Moreover, the 
direction of the shear band is either horizontal or vertical because of the isotropy of the material.
Figure~\ref{fig:histgram}(a) shows the histogram of the location of the shear band for the two-dimensional stable glass  with $T_{\rm ini}=0.035$ and $N=64000$ 
($L=253$). The abscissa denotes the $x$-coordinate of the vertical shear band or the $y$-coordinate of the horizontal shear band. The number of independent 
realizations for this measurement is $100$.
The histogram for the homogenous samples is relatively uniform, confirming that the shear-band location is random in space.  We also measure the fraction of 
the samples having the horizontal shear band, $f_{\rm H}$. We find $f_{\rm H}\approx 0.45$ for the homogenous case (no seed), which implies that the horizontal and 
vertical shear bands happen with essentially the same probability.

We now examine the effects of the seed in terms of location and  direction of the shear band separately.
First, Fig.~\ref{fig:histgram}(c) shows the result for a small circular seed with diameter $D=8$ (or $D_a=D_b=8$). 
The histogram has a tiny peak at the center, which means that the probability of finding the shear band at the position of the seed is higher than 
in the homogeneous sample case.  As the size of the circular seed $D$ is increased, the peak is enhanced, and most of the samples have the shear band 
appearing at the center (see Fig.~\ref{fig:histgram}(d)). However, $f_{\rm H}$ is still near $0.5$ for both $D=8$ and $D=16$, as expected from the symmetry 
of the circular seed.

We now insert an elliptical seed. We vary $D_a$ while we fix $D_b=8$.
We find that the peak at the center is defined more sharply with increasing $D_a$, as shown in Figs.~\ref{fig:histgram}(e,f). Note that the area of the circular seed in Fig.~\ref{fig:histgram}(d) and of the elliptical seed in Fig.~\ref{fig:histgram}(f) are the same. Moreover, $f_{\rm H}$ quickly departs from near $\sim 0.5$ once the 
elliptical seed is introduced, which means that, now, the direction of the shear band is statistically controlled by the orientation of the seed. This is consistent with 
the observation that the shear band tends to take place where the Eshelby events are aligned.

\section{Propagation of Eshelby events}

\label{app:propagation}

\begin{figure}
\includegraphics[width=0.48\columnwidth]{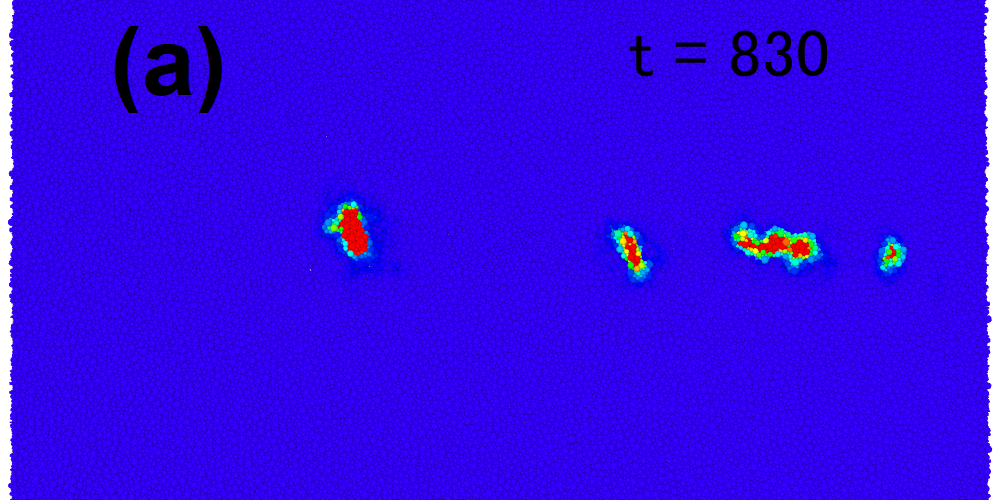}
\includegraphics[width=0.48\columnwidth]{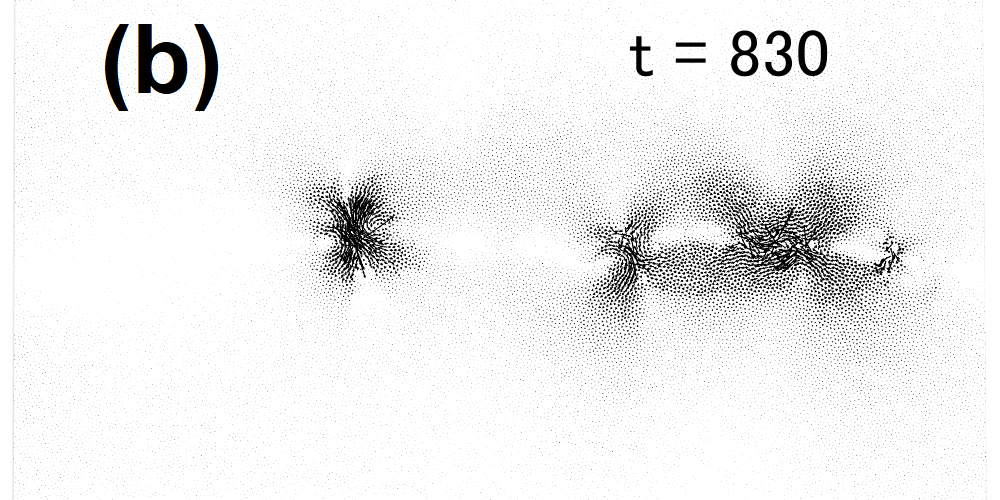}
\includegraphics[width=0.48\columnwidth]{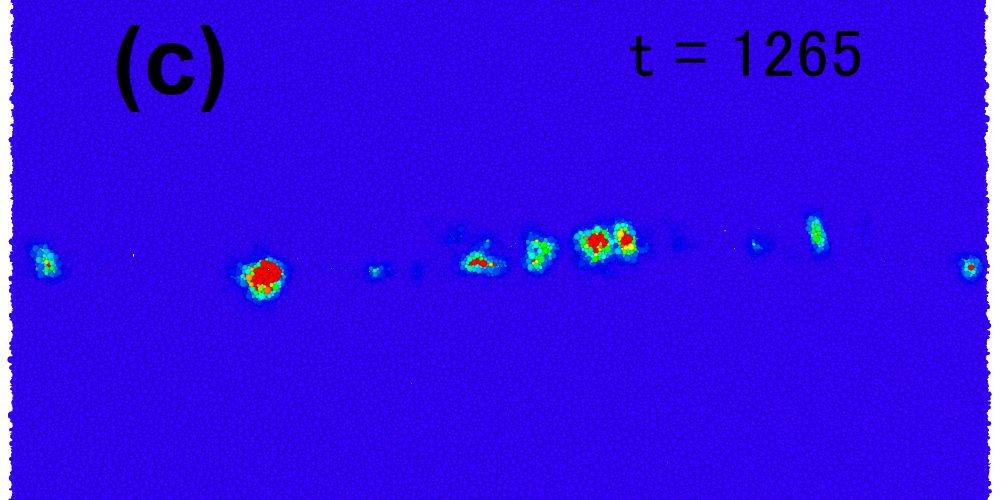}
\includegraphics[width=0.48\columnwidth]{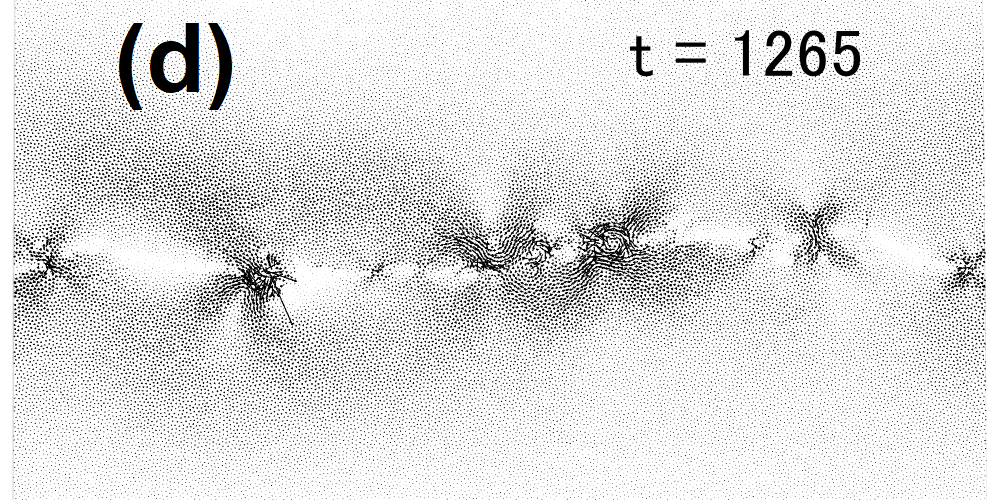}
\includegraphics[width=0.48\columnwidth]{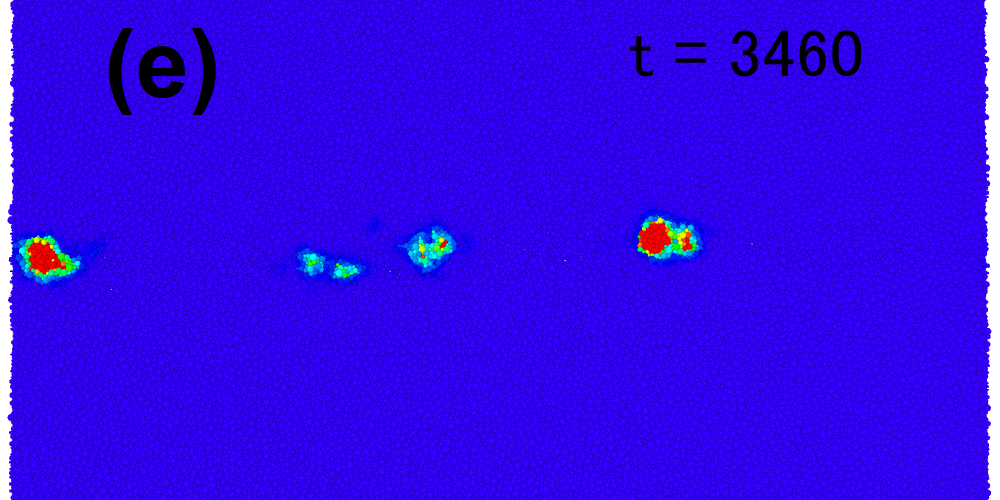}
\includegraphics[width=0.48\columnwidth]{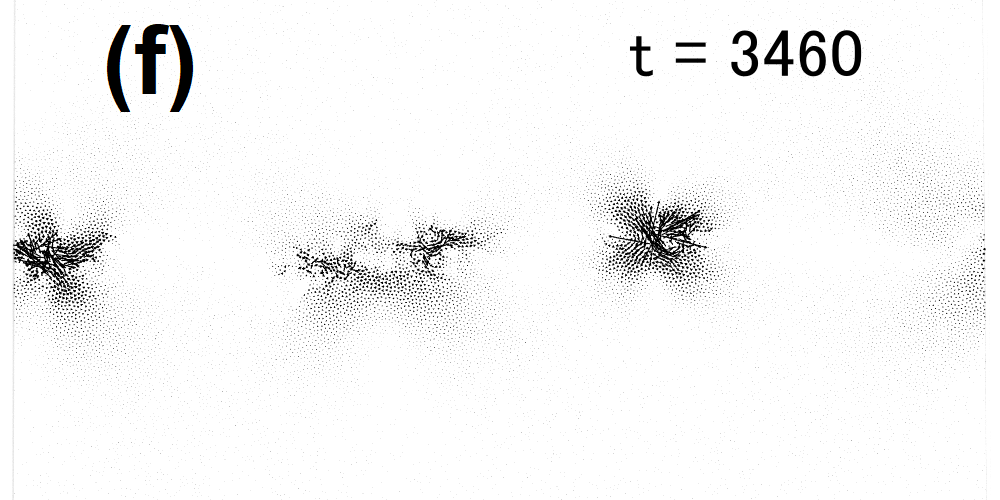}
\caption{
Eshelby propagation during the gradient-descent dynamics for a stable glass with $T_{\rm ini}=0.035$ and $D_a=32$. Left (a,c,e): $D_{\rm min}^2(t-\Delta t, t)$ 
between successive time frames. We set $\Delta t=5$.
Right (b,d,f): The corresponding displacement vectors of the left panels. The length of the arrows is magnified by a factor 30.  
}
\label{fig:during_yielding2}
\end{figure}  

In Fig.~\ref{fig:during_yielding} in the main text, we see the intermittent behavior in the mean squared velocity, $v^2$, during the gradient-descent dynamics starting from the configuration 
right before the largest stress drop for a stable glass ($N=64000$, $T_{\rm ini}=0.035$, and $D_a=32$).
To see this intermittent behavior in real space, we visualize the displacement vector fields between two successive time frames in Fig.~\ref{fig:during_yielding2}.
We observe that Eshelby-like displacements propagate along the shear band. Remarkably, the Eshelby vector fields propagate many times during the entire lapse of 
shear band formation. This observation appears reasonable because a single Eshelby displacement vector can carry only a tiny displacement at its core. Thus, to build 
a system-spanning macroscopic shear band, the Eshelby displacement vectors have to propagate many times, as theoretically argued~\cite{dasgupta2013yield}.

\section{Stress inside the seed}

\label{app:stress}

\begin{figure}
\includegraphics[width=0.95\columnwidth]{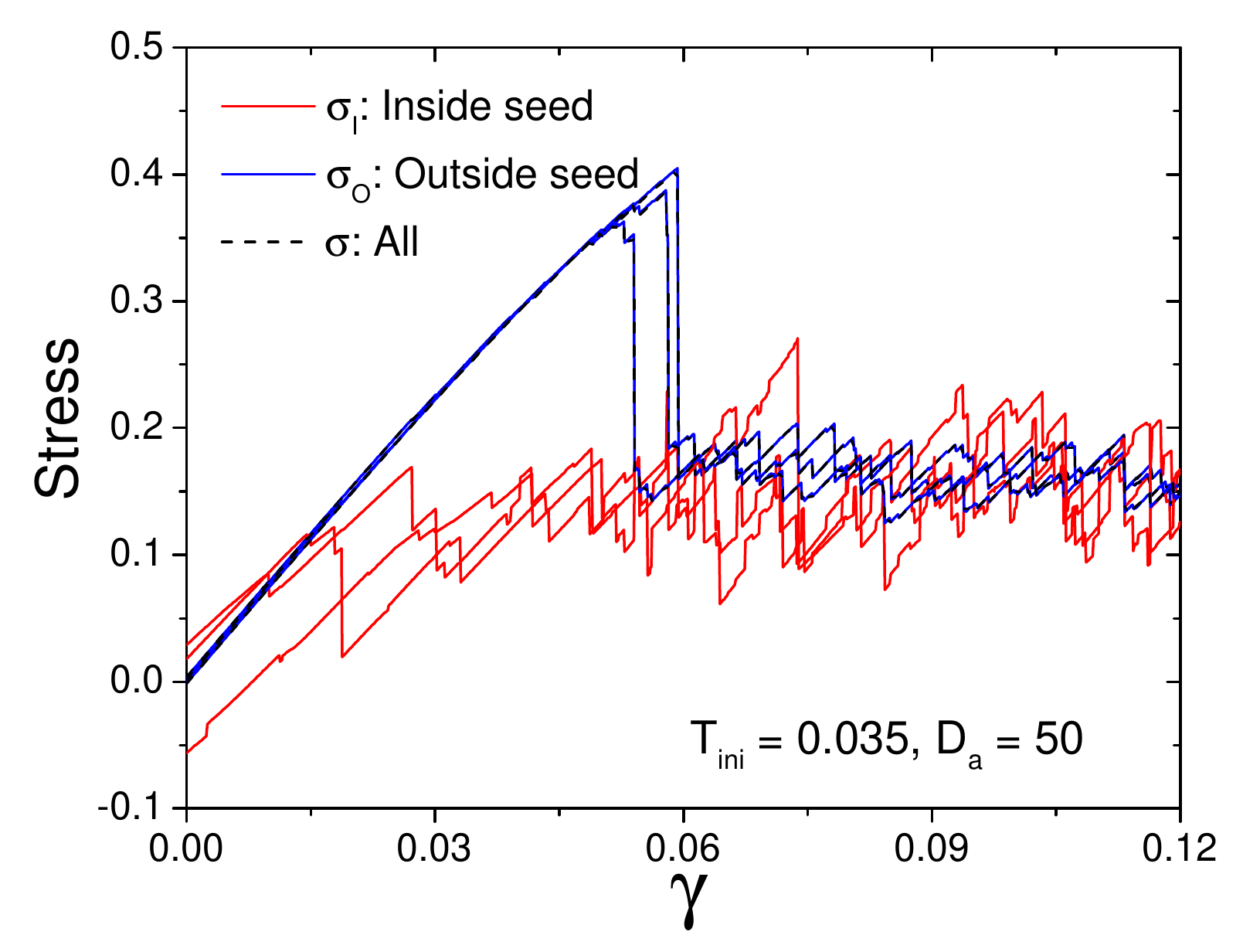}
\caption{Stress inside and outside the seed, as well as for the entire system, for stable glass samples with $N=64000$, $T_{\rm ini}=0.035$ and $D_a=50$.
Three individual realizations are shown.}
\label{fig:stress_seed}
\end{figure}

We monitor the stress evolution under external strain inside and outside the seed separately. 

First, we introduce the stress inside and outside the seed.
We rewrite Eq.~(\ref{eq:shear_stress}) as 
\begin{equation}
\sigma = \frac{1}{N} \sum_i \sigma_i,
\end{equation}
where $\sigma_i$ is defined by
\begin{equation}
\sigma_i = \frac{1}{2} \sum_{j \neq i} \frac{x_{ij}y_{ij}}{r_{ij}} v_{ij}'(r_{ij}).
\end{equation}
We define the stress inside the seed, $\sigma_{\rm I}$, and outside the seed, $\sigma_{\rm O}$, by
\begin{eqnarray}
\sigma_{\rm I} &=& \frac{1}{N_{\rm I}} \sum_{i \in {\rm I}} \sigma_i, \\
\sigma_{\rm O} &=& \frac{1}{N_{\rm O}} \sum_{i \in {\rm O}} \sigma_i,
\end{eqnarray}
where $N_{\rm I}$ and $N_{\rm O}$ are the number of particles inside and outside the seed, respectively ($N=N_{\rm I}+N_{\rm O}$).
Oviously, one has 
$\sigma=c_{\rm I} \sigma_{\rm I}+c_{\rm O} \sigma_{\rm O}$, where $c_{\rm I}=N_{\rm I}/N$ and $c_{\rm O}=N_{\rm O}/N$.

Figure~\ref{fig:stress_seed} shows the stress inside and outside the seed and for the entire sample as a function of the strain $\gamma$ for a stable glass with a seed of size $D_a=50$. Three independent samples are shown. The stress inside the seed shows multiple drops much before the macroscopic yield strain, $\gamma_{\rm Y} \simeq 0.06$, 
and displays ductile-like yielding behavior. On the other hand, the stress outside the seed shows elastic response, having an essentially identical curve to that of the entire 
system. The number of particles inside the seed with $D_a=50$ is around $N_{\rm I} \approx 300$, and hence $c_{\rm I} \approx 0.005$. Thus, in terms of the stress value, 
the seed  contributes only marginally to the entire sample. Yet, as demonstrated in Fig.~\ref{fig:stress} in the main text, the location of $\gamma_{\rm Y}$ is significantly affected by the 
presence of the seed. This observation confirms that a rare soft region with a concentration that vanishes in the thermodynamic limit can impact the macroscopic yielding 
behavior.

\section{Local stress}

\label{app:local}

\begin{figure}
\includegraphics[width=0.48\columnwidth]{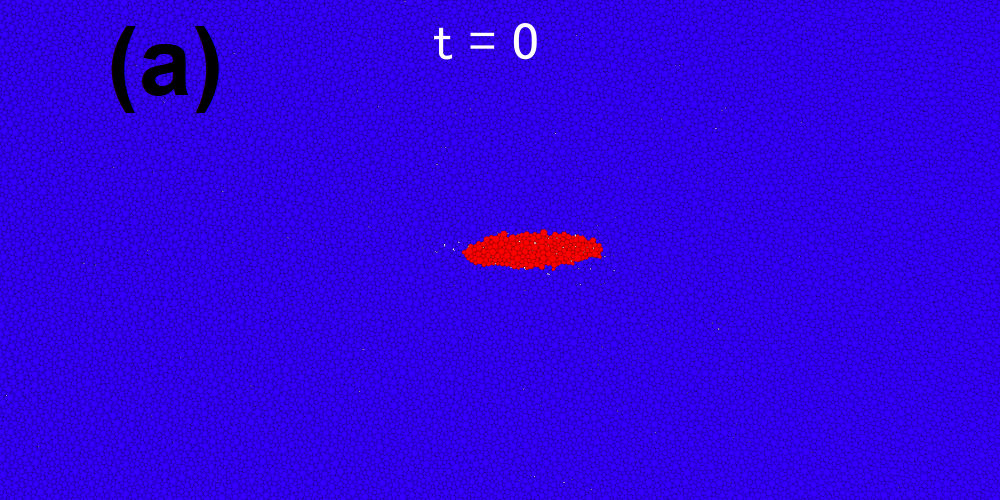}
\includegraphics[width=0.48\columnwidth]{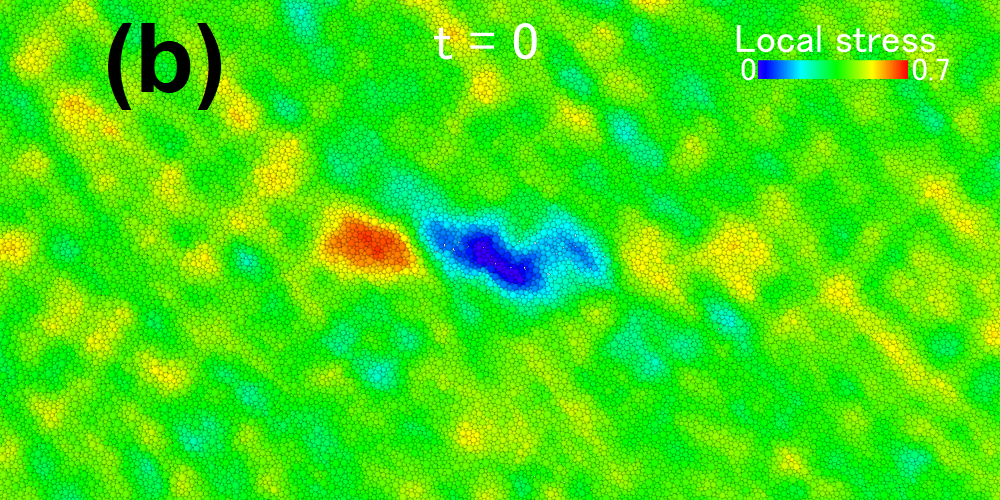}
\includegraphics[width=0.48\columnwidth]{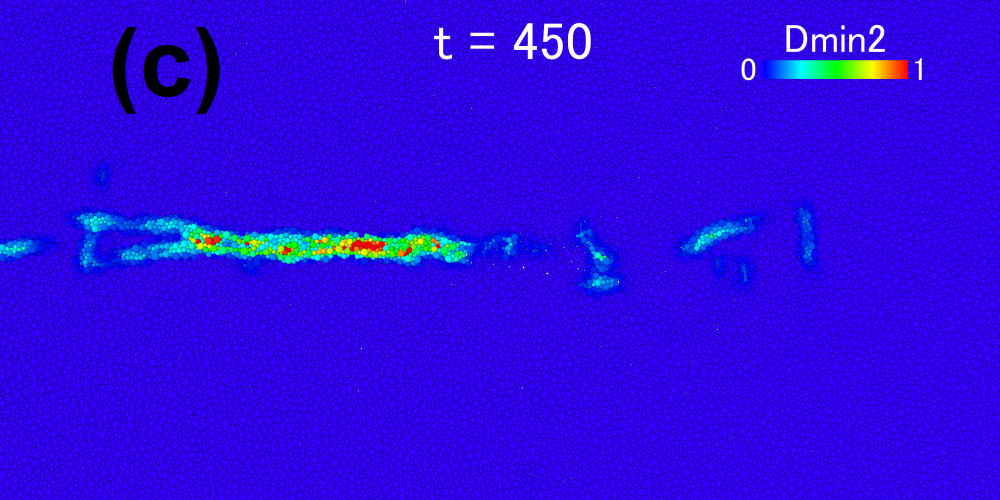}
\includegraphics[width=0.48\columnwidth]{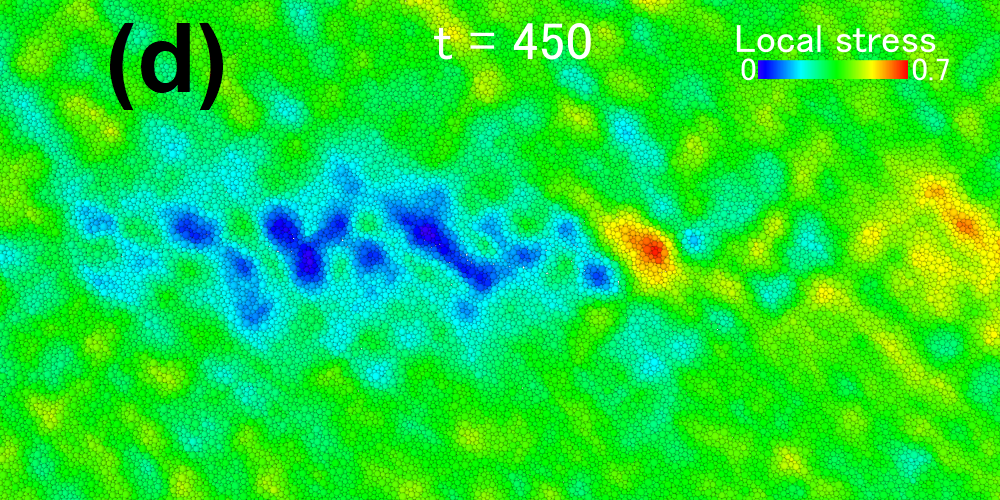}
\includegraphics[width=0.48\columnwidth]{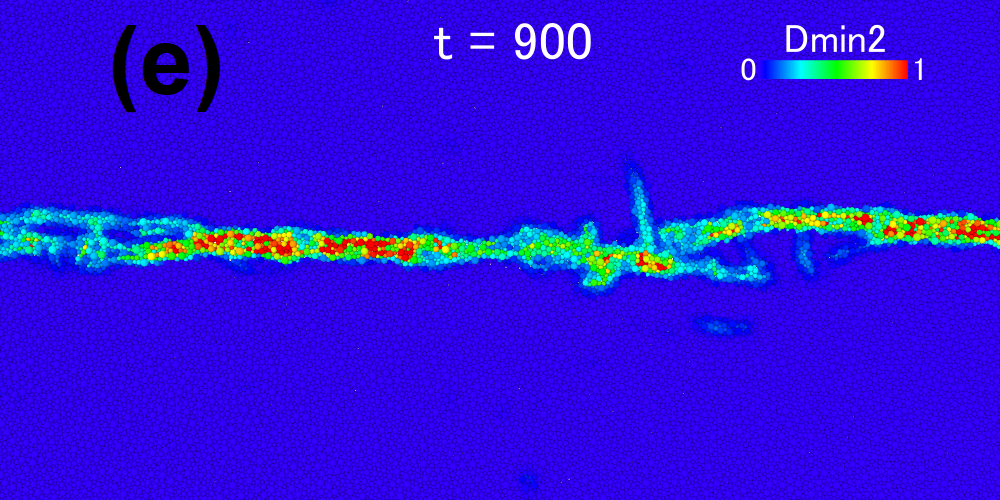}
\includegraphics[width=0.48\columnwidth]{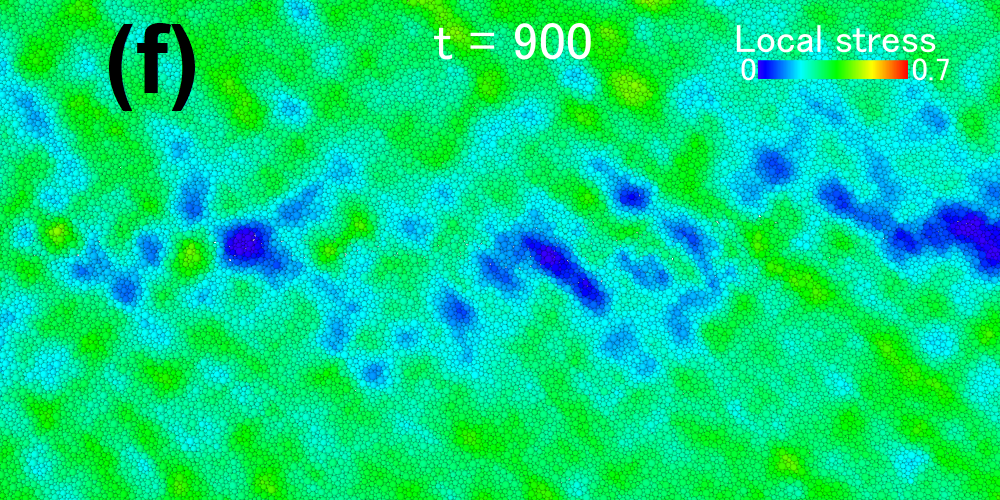}
\caption{
Snapshots of the gradient-descent dynamics for a macroscopic brittle yielding. The snapshots are obtained from a sample with $N=64000$, $T_{\rm ini}=0.035$, and 
$D_a=32$.
Left (a,c,e): Nonaffine displacement field, $D_{\rm min}^2$, between $t=0$ and $t$. Right (b,d,f): Local stresses corresponding to the left panels. }
\label{fig:local_stress}
\end{figure}  

We investigate the local stresses near the seed at the initial stage of shear-band formation. 
We define the local stress $\sigma_i^{\rm local}$ by
\begin{equation}
    \sigma_i^{\rm local} = \frac{1}{n_i} \sum_{\substack{j \\ (r_{ij} < R)}} \sigma_j,
\end{equation}
where $n_i$ is the number of neighboring particles for the $i$th particle. The local stress is thus obtained by averaging over particles within a cut-off radius $R$. We set $R=5$, 
following Ref.~\cite{barbot2018local}.

Figure~\ref{fig:local_stress} shows the time evolution of the nonaffine displacement $D_{\rm min}^2$ (left panels) and of the local stress (right panels) during the 
gradient-descent dynamics for the largest stress drop. At $t=0$, i.e., right before the macroscopic yielding, the entire sample is stressed rather homogeneously except 
near the seed (Fig.~\ref{fig:local_stress}(b)).
Because the seed region has already yielded along the quasi-elastic branch, it has a lower stress, which appears in blue. This inhomogeneity of the local stress at the 
seed induces higher stresses near the tips of the seed, yet with a magnitude of order 1. 
The higher stress field near the tips then induces plastic activity at later time (Figs.~\ref{fig:local_stress}(c,d)), eventually causing propagation of the shear band (Figs.~\ref{fig:local_stress}(e,f)).  

\section{Finite-size effects in a well-annealed glass}

\label{app:finite}

\begin{figure}
\includegraphics[width=0.95\columnwidth]{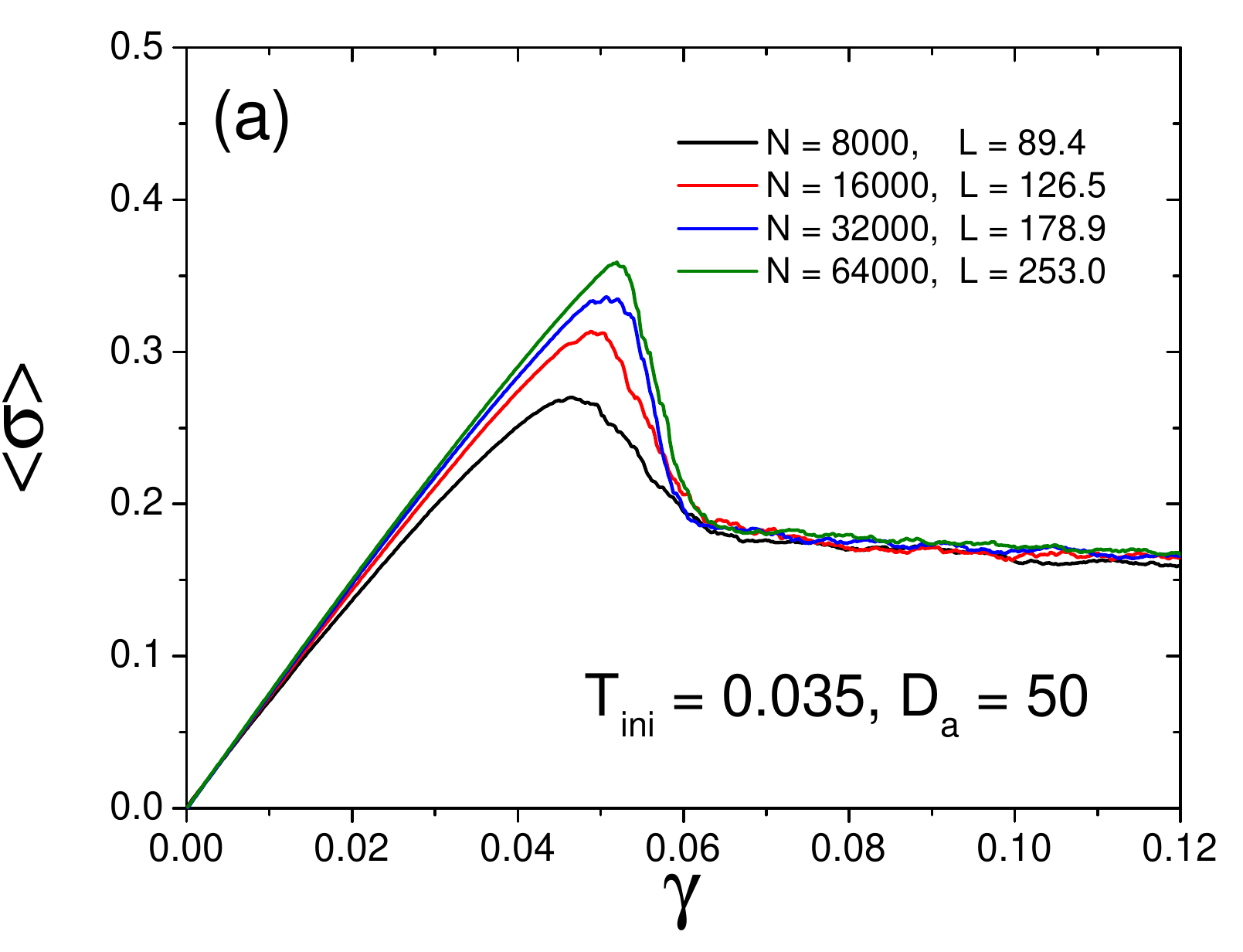}
\includegraphics[width=0.95\columnwidth]{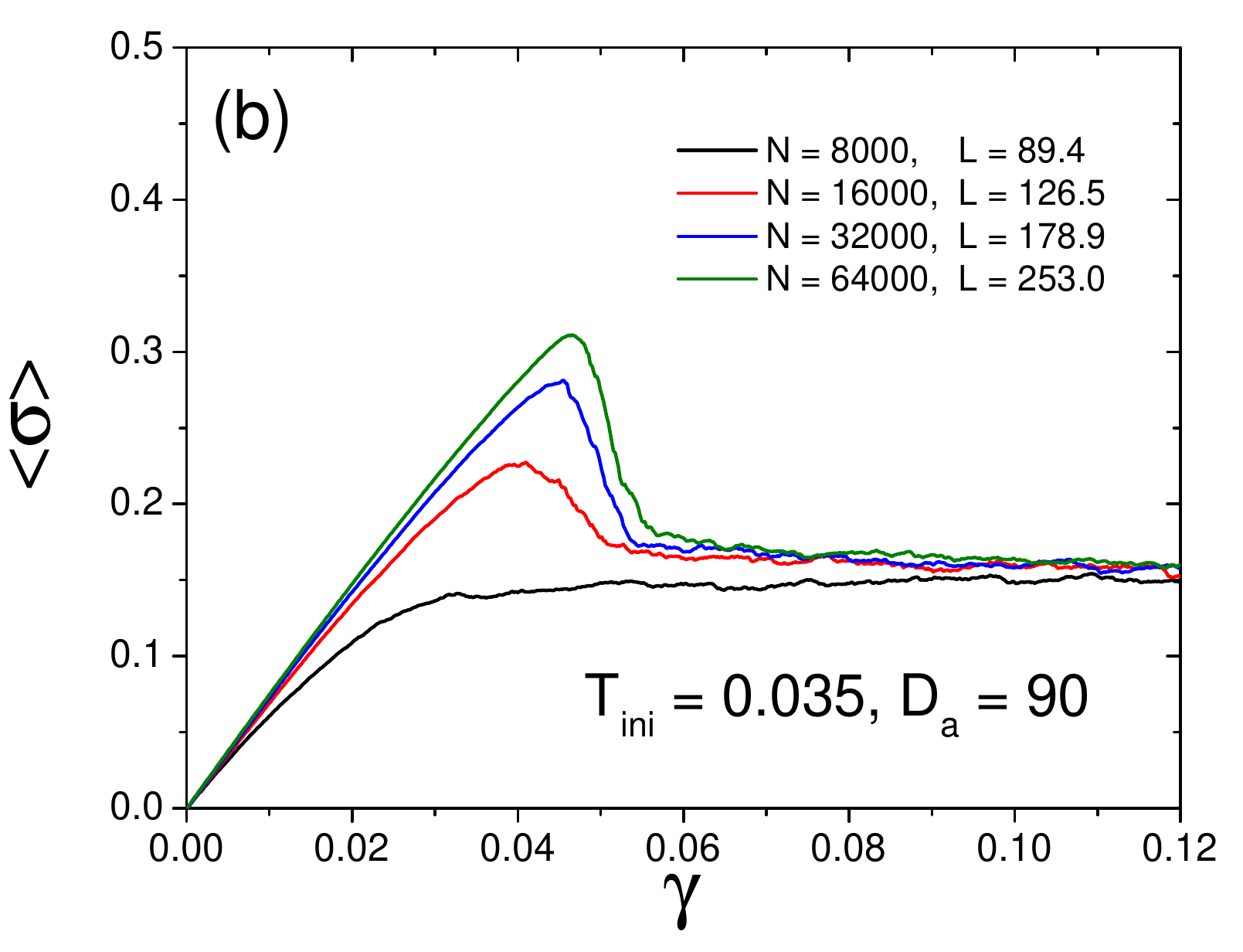}
\caption{Averaged stress versus strain curves of a stable glass ($T_{\rm{ini}}=0.035$) for fixed $D_a=50$ (a) and $D_a=90$ (b) for several system sizes $N$ (and 
corresponding linear box length $L=N^{1/2}$ for a density $\rho=1$).
}
\label{fig:varying_N}
\end{figure} 

We consider the variation of the yielding behavior with the system size $L$ when the size of the seed is fixed. We have argued in the main text that the appropriate 
thermodynamic limit is to consider $L\to \infty$ first, and only then  $D_a\to \infty$. This avoids considering a seed that scales as the system size or, within a finite system 
with periodic boundary conditions, a seed that is influenced by its images. We illustrate this points by showing in Fig.~\ref{fig:varying_N} the finite-size effect on the 
stress versus strain curves for fixed $D_a$ (and $D_b$) and for a well-annealed glass with $T_{\rm{ini}}=0.035$.

In Fig.~\ref{fig:varying_N}(a), we consider a seed length $D_a=50$ that is about half of the linear box length of the smallest system under study, i.e., $N=8000$ and we 
investigate how the curve evolves when increasing $N$ up to $64000$. All system sizes show a qualitatively similar behavior with a large stress overshoot. The slope of the 
drop after the overshoot increases with $N$, which is in agreement with the finite-size scaling of a discontinuous (brittle) yielding transition as already found in the absence 
of a seed. Quite notably, the overshoot becomes more prominent as the system size increases, all in all confirming the brittle nature of the transition in the thermodynamic limit.  

On the other hand, in Fig.~\ref{fig:varying_N}(b), we set $D_a=90$, so that the seed now spans the entire sample for the smallest system of $N=8000$. In this latter case, 
the stress overshoot is completely wiped out and a seemingly ductile behavior is observed. However, when increasing the system size the overshoot reappears and becomes 
more and more prominent with, as for $D_a=50$, the slope of the curve that becomes steeper as $N$ increases. This clearly indicates a brittle yielding in the 
thermodynamic limit.


\section{Variation of $D_b$}

\label{app:db}

\begin{figure*}[htbp]
\includegraphics[width=0.66\columnwidth]{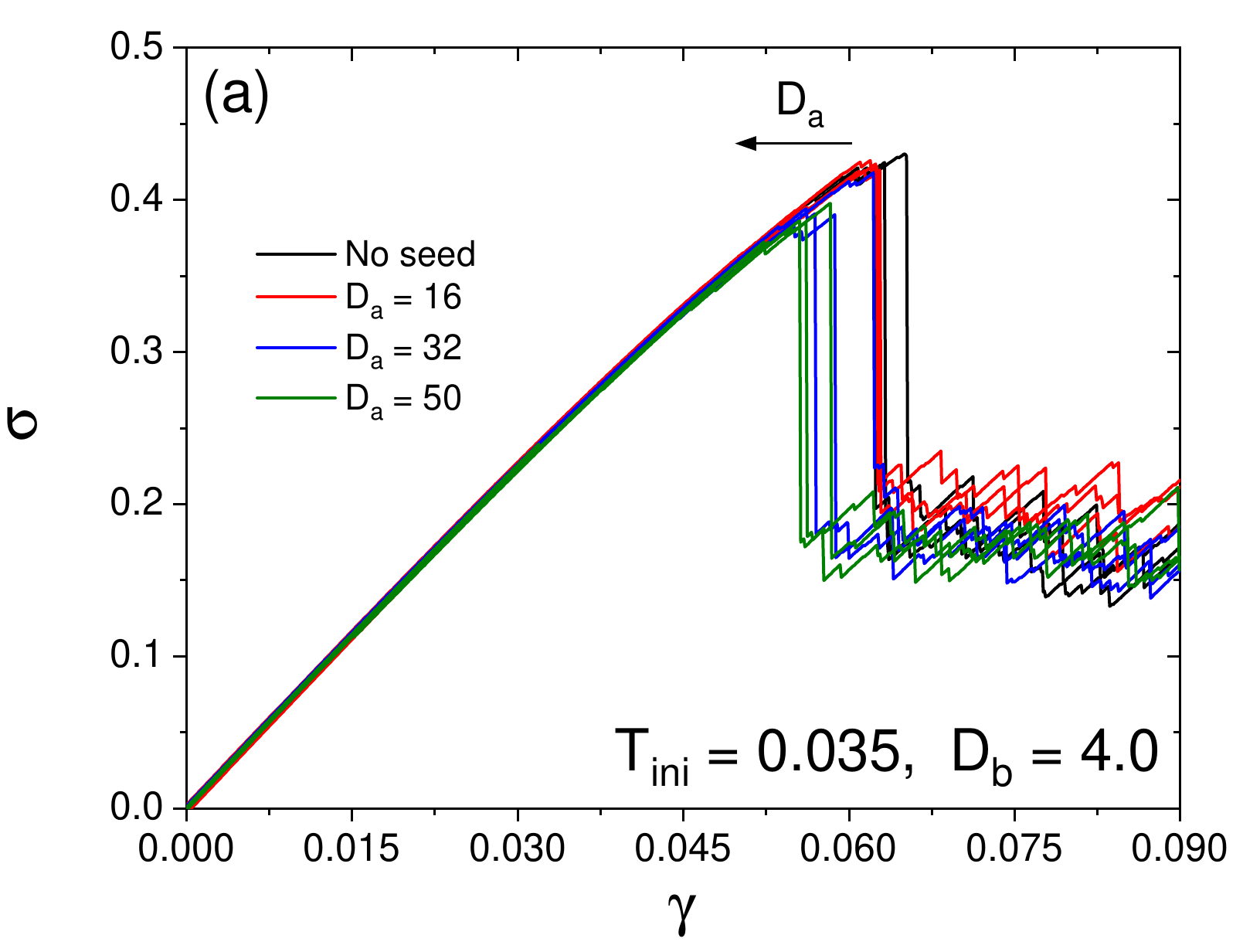}
\includegraphics[width=0.66\columnwidth]{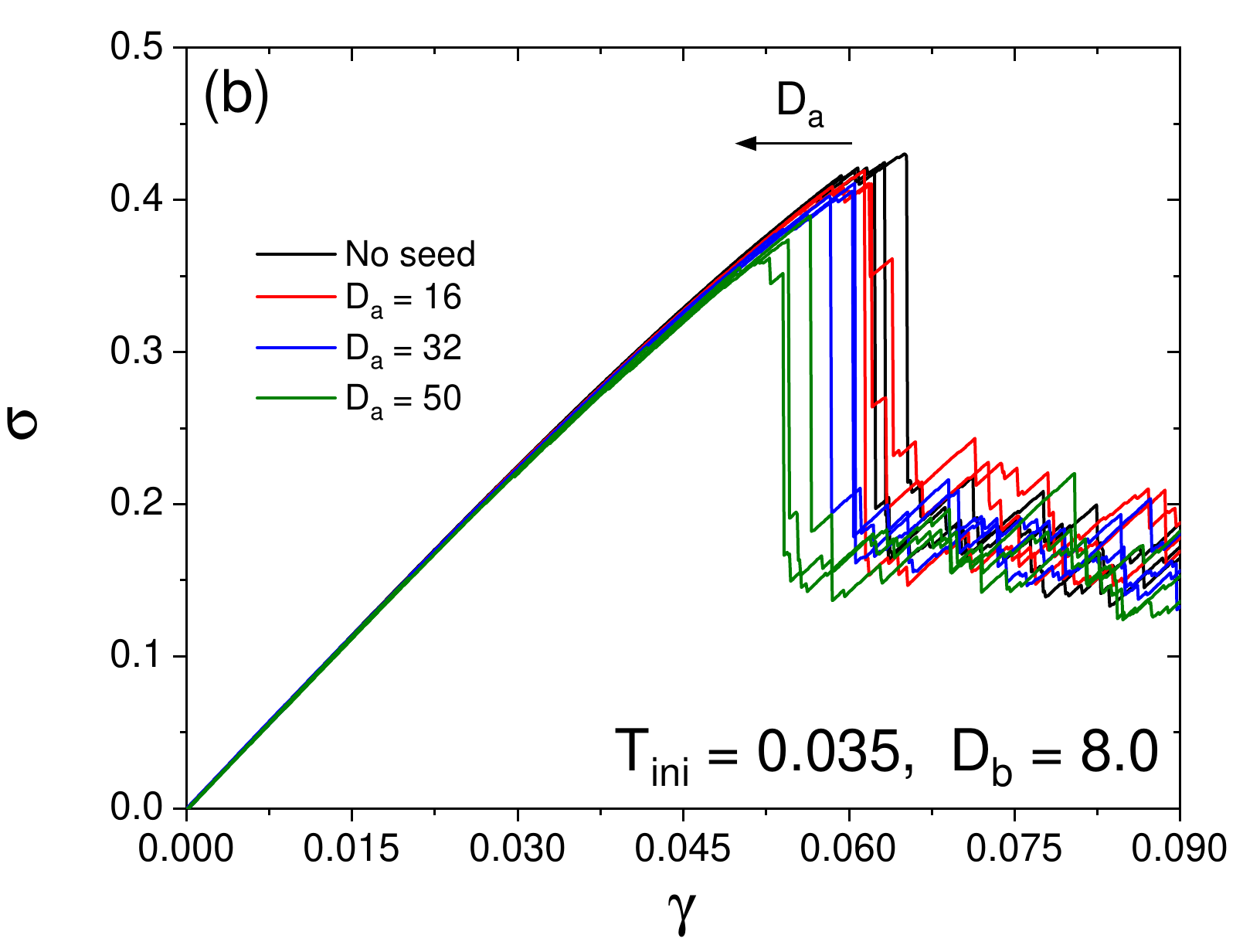}
\includegraphics[width=0.66\columnwidth]{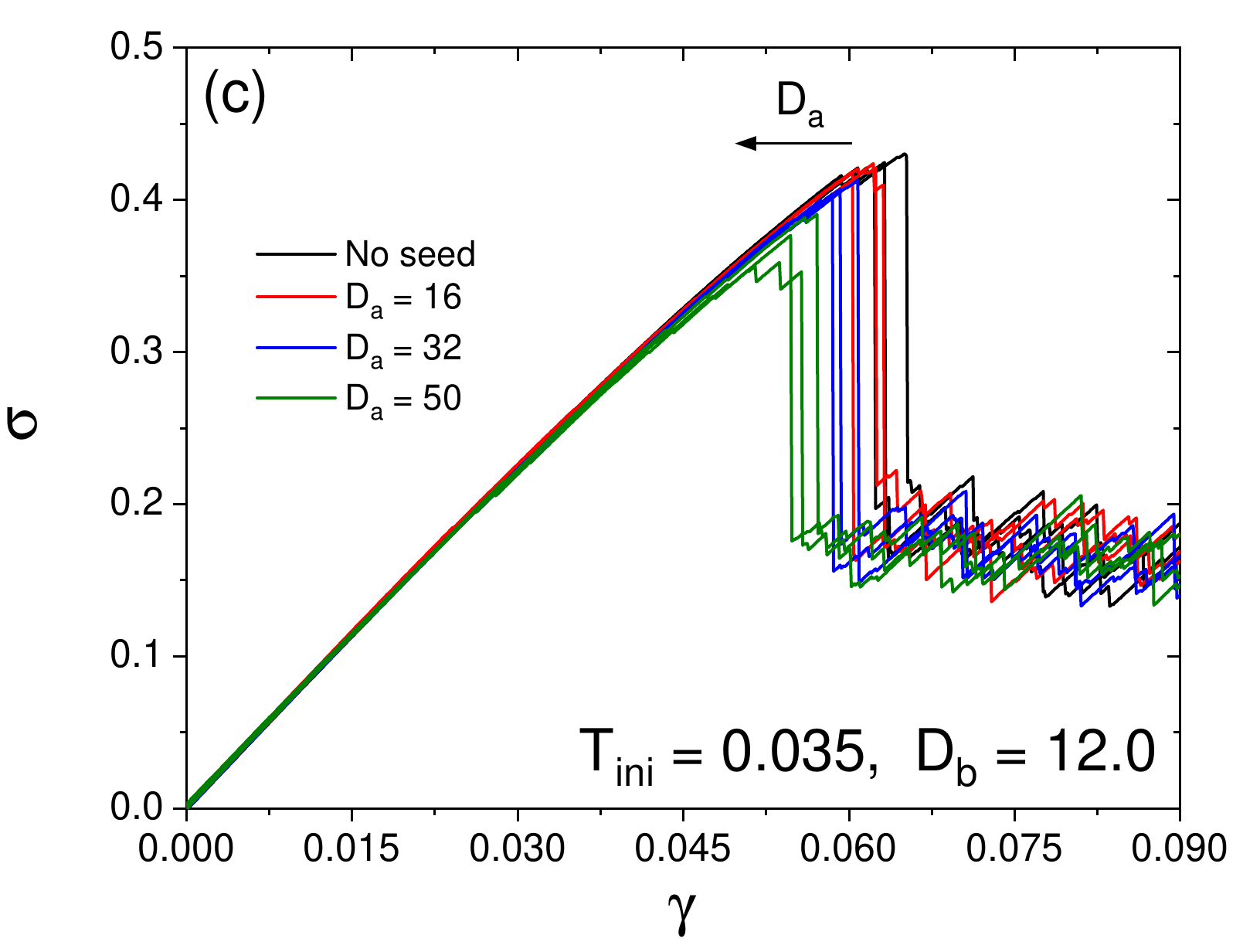}
\includegraphics[width=0.66\columnwidth]{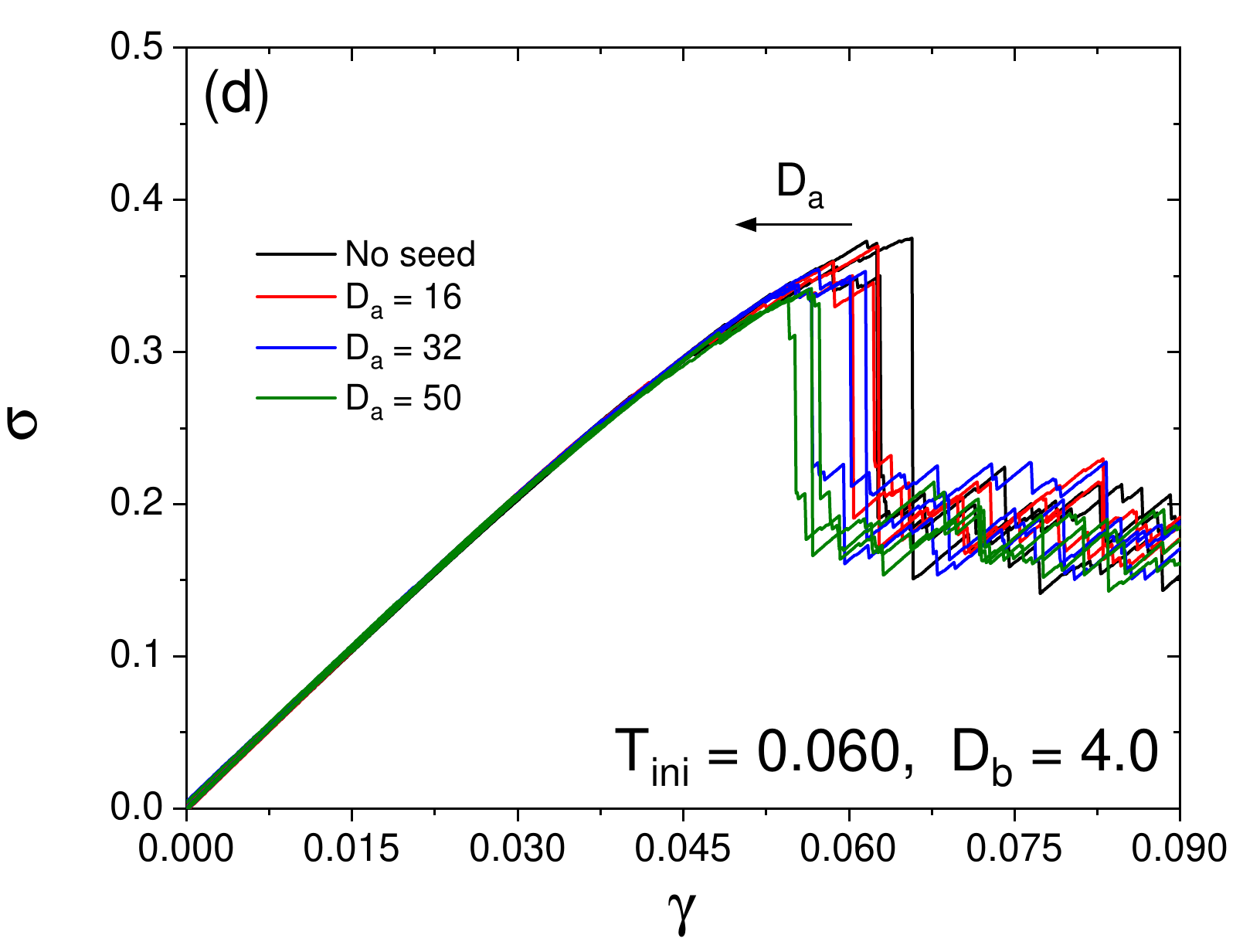}
\includegraphics[width=0.66\columnwidth]{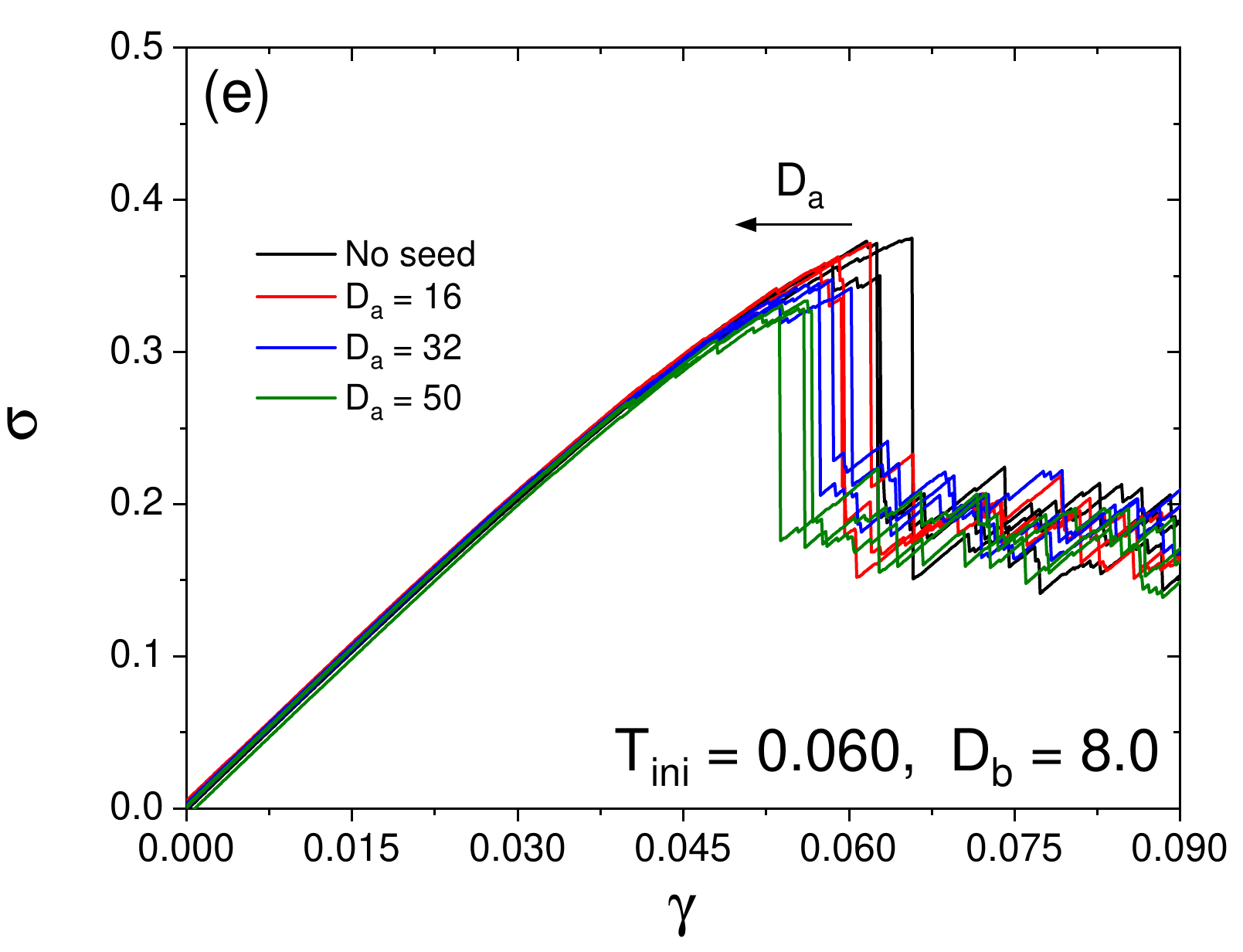}
\includegraphics[width=0.66\columnwidth]{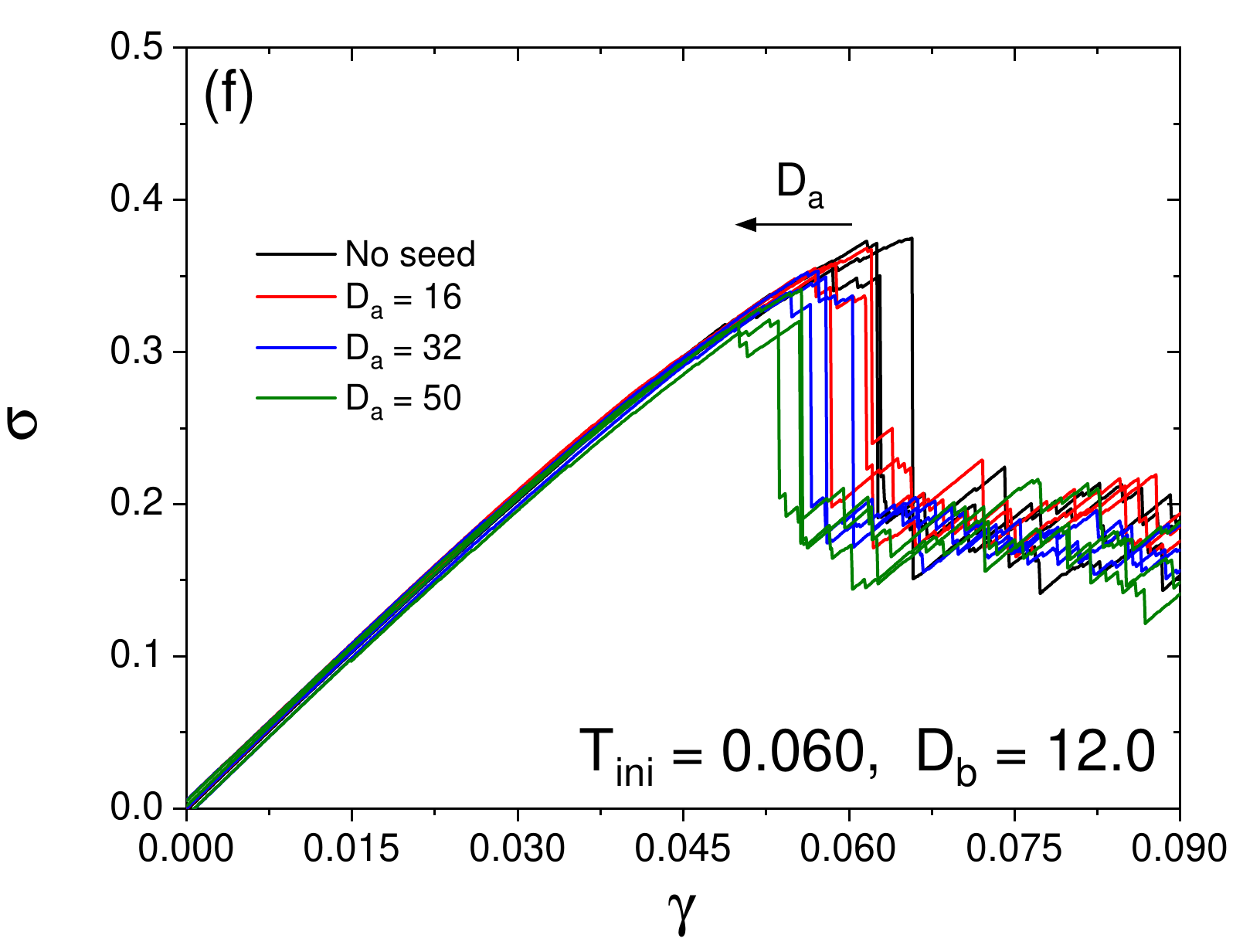}
\caption{The stress versus strain curves of $2d$ glass samples with $N=64000$ atoms for various $D_b$.
Three independent realisations for each $D_a$ are shown for glasses with $T_{\rm ini}=0.035$ (a,b,c) and $T_{\rm ini}=0.060$ (d,e,f).}
\label{fig:variation_D_b}
\end{figure*}

In the main text, we choose $D_b=8$
which is of the order of the typical scale of an elementary rearranging region~\cite{barbot2018local}.
However, this scale might change with the degree of annealing, as the typical scale of the quasi-localized defect changes with parent temperature~\cite{rainone2020pinching}. 
In Fig.~\ref{fig:variation_D_b}, we show the stress versus strain curves of $2d$ glass samples with varying both $D_a$ and $D_b$ for $T_{\rm ini}=0.035$ and $T_{\rm ini}=0.060$ (both are in the brittle regime).
We observe qualitatively the same behavior as FIG. 1 of the main text, justifying our choice, $D_b=8$.

\bibliography{yielding.bib}

\end{document}